\documentclass[traditabstract]{aa}

\newcommand{\lsim}{\lower 2pt \hbox{$\, \buildrel {\scriptstyle
<}\over {\scriptstyle \sim}\,$}}  \newcommand{\gsim}{\lower 2pt
\hbox{$\, \buildrel {\scriptstyle >}\over {\scriptstyle \sim}\,$}}
\usepackage{appendix}
\usepackage{amsmath}
\usepackage{graphicx}
\usepackage{txfonts}
\usepackage{natbib}
\usepackage{longtable}
\usepackage{rotating}
\usepackage{lscape}
\usepackage{float}
\usepackage{subfig}
\usepackage{multirow}
\usepackage{array}
\usepackage{fixltx2e}
\usepackage{hyperref}

\bibpunct{(}{)}{;}{a}{}{,}

\newcommand{\Pal}{Pa$\alpha$ at 1.876\,${\mu}$m}
\newcommand{\Brdl}{Br$\delta$ at 1.945\,${\mu}$m}
\newcommand{\Brgl}{Br$\gamma$ at 2.166\,${\mu}$m}

\newcommand{\Pa}{Pa$\alpha$}
\newcommand{\Brd}{Br$\delta$}
\newcommand{\Brg}{Br$\gamma$}

\newcommand{\LTIR}{L$_{\rm IR}$}
\newcommand{\Lsolar}{L$_{\odot}$}
\newcommand{\Av}{A$_{\rm V}$}
\newcommand{\Reff}{R$_{\rm eff}$}
\newcommand{\Si}{$\Sigma_{\rm SFR}$}

\newcommand\nodata{ ~$\cdots$~ }

\newcounter{subfig}
\newcounter{fake_fig}

\newcolumntype{x}[1]{%
>{\raggedleft\hspace{0pt}}p{#1}}%

\newcolumntype{z}[1]{%
>{\raggedright\hspace{0pt}}p{#1}}%

\begin{document}

\title{VLT-SINFONI integral field spectroscopy of low-z luminous and ultraluminous infrared galaxies}

\subtitle{II. 2D extinction structure and distance effects}

\author{J. Piqueras L\'opez\inst{1}
\and L. Colina\inst{1}
\and S. Arribas\inst{1}
\and A. Alonso-Herrero\inst{2}
}
\offprints{Javier Piqueras L\'opez\\ {\tt piqueraslj@cab.inta-csic.es} \smallskip}

\institute{
Centro de Astrobiolog\'ia (INTA-CSIC), Ctra de Torrej\'on a Ajalvir, km 4, 28850, Torrej\'on de Ardoz, Madrid, Spain
\and Instituto de F\'isica de Cantabria, CSIC-UC, Avenida de los Castros S/N, 39005 Santander, Spain
}

\date{  }

\abstract{We present a 2D study of the internal extinction on (sub)kiloparsec scales of a sample of local ($z<0.1$) LIRGs (10) and ULIRGs (7), based on near-infrared \Pa, \Brd, and \Brg\ line ratios, obtained with VLT-SINFONI integral-field spectroscopy (IFS). The 2D extinction (\Av) distributions of the objects, map regions of $\sim3\times3$\,kpc (LIRGs) and $\sim12\times12$\,kpc (ULIRGs), with average angular resolutions (FWHM) of $\sim$0.2\,kpc and $\sim$0.9\,kpc, respectively. The individual \Av\ galaxy distributions indicate a very clumpy dust structure already on sub-kiloparsec scales, with values (per spaxel) ranging from \Av$\sim$1 to 20\,mag in LIRGs, and from \Av$\sim$2 to 15\,mag in ULIRGs. As a class, the median values of the distributions are \Av=5.3\,mag and \Av=6.5\,mag for the LIRG and ULIRG subsamples, respectively. In $\sim$70\% of the objects, the extinction peaks at the nucleus with values ranging from \Av$\sim$3 to 17\,mag. Within each galaxy, the \Av\ radial profile shows a mild decrement in LIRGs within the inner 2\,kpc radius, while the same radial variation is not detected in ULIRGs, likely because of the lower linear scale resolution of the observations at the distance of ULIRGs. 
We evaluated the effects of the galaxy distance in the measurements of the extinction as a function of the linear scale (in kpc) of the spaxel (i.e. due to the limited angular resolution  of the observations). If the distribution of the gas/dust and star-forming regions in local LIRGs (63\,Mpc, 40\,pc/spaxel on average) is the same for galaxies at greater distances, the observed median \Av\ values based on emission line ratios would be a factor $\sim$ 0.8 lower at the average distance of our ULIRG sample (328\,Mpc, 0.2\,kpc/spaxel), and a factor $\sim$0.67 for galaxies located at distances of more than 800\,Mpc (0.4\,kpc/spaxel). This distance effect would have implications for deriving the intrinsic extinction in high-z star-forming galaxies and for subsequent properties such as star formation rate, star formation surface density, and KS- law, based on H$\alpha$ line fluxes. If local LIRGs are analogues of the main-sequence (MS) star-forming galaxies at cosmological distances, the extinction values (\Av) derived from the observed emission lines in these high-z sources would need to be increased by a factor 1.4 on average.}


\authorrunning{J. Piqueras L\'opez et al.}
\titlerunning{VLT-SINFONI IFS of low-z Luminous and Ultraluminous Infrared Galaxies II.}
\maketitle 
%
\setcounter{figure}{2}
\section{Introduction}

Since the first results obtained by the  \emph{Infrared Astronomical Satellite} (\emph{IRAS}) \citep{Soifer:1984p8204}, there has been strong effort to study the physical processes that power the luminous (LIRGs; $10^{11}$\Lsolar$<$\LTIR$< 10^{12}$\Lsolar) and ultraluminous (ULIRGs; $10^{12}$\Lsolar$<$\LTIR$< 10^{13}$\Lsolar) infrared galaxy population (\citealt{Sanders:1996p845}, \citealt{Lonsdale:2006p4228}). The origin of the mid- and far-infrared emission (\LTIR[8-1000\,$\mu$m]) that dominates their bolometric luminosity is established as mainly due to massive starbursts with a small AGN contribution for LIRGs and with an increasing contribution in ULIRGs (e.g. \citealt{1995ApJ...444...97G}, \citealt{Veilleux:2009gy}, \citealt{Nardini:2010p405}, \citealt{Alonso-Herrero:2012p744} and references therein). The radiation that originates in the starburst and/or the active galactic nucleus is then reprocessed by a surrounding dust component, and then re-emitted at long wavelengths in the form of a huge infrared emission.

One of the main difficulties in understanding the underlying power source of the LIRGs and ULIRGs is the high opacity of their nuclear regions. Previous optical \citep{GarciaMarin:2009p8459} and near-infrared studies in LIRGs and ULIRGs (\citealt{1998ApJ...498..579G}, \citealt{Scoville:2000AJ119}, \citealt{AlonsoHerrero:2006p4703}) reveal that the distribution of the dust in these object is not uniform and that, though the dust tends to concentrate in the inner kiloparsecs with average visual extinction of \Av$\sim$3-5\,mag in LIRGs and even higher in ULIRGs, the global distribution shows a patchy structure on kiloparsec and sub-kiloparsec scales (\citealt{Colina:2000ApJ533}, \citealt{GarciaMarin:2006ApJ650}, \citealt{Bedregal:2009p2426}).

Besides the importance of knowing the 2D structure of the dust to understand the environment where the power source of the LIRGs and ULIRGs is embedded, dust plays a key role in the derivation of other physical and structural parameters of these objects, such as the derived star formation rate \citep{GarciaMarin:2009p8459}, the effective radius \citep{Arribas:2012p1203}, and as a consequence, the dynamical masses. 

Understanding the distribution and effect of dust in star-forming galaxies is also important for correctly interpreting or comparing different tracers of star formation during the history of the Universe. This is in turn relevant when comparing local and high-z star-forming galaxy populations, which are often observed using different tracers and/or resolutions. The distribution of dust can in principle be studied in detail in local U/LIRGs with the advantage of the relatively high linear resolution and S/N. These studies can, therefore, help us interpret observations of analogous high-z star-forming galaxies, for which such a level of resolution, and S/N is not attainable with current instruments.

The present work is part of a series presenting new H- and K-band SINFONI (\emph{Spectrograph for INtegral Field Observations in the Near Infrared}, \citealt{Eisenhauer:2003p8484}) seeing-limited observations of a sample of local LIRGs and ULIRGs. \cite{Piqueras2012A&A546A} (hereafter Paper I) presented the atlas of the sample, the data reduction, and a brief analysis and discussion of the morphology of the gas emission and kinematics. In this second paper in the series, we focus on the study of the 2D distribution of the dust derived using the \Brg/\Brd\ and \Pa/\Brg\ ratios for LIRGs and ULIRGs, respectively, whereas in Piqueras L\'opez et al. 2013 (Paper III, in preparation), we will apply the results for the 2D dust structure to study both the overall star formation rate (SFR) and the kpc structure of the SFR surface density ($\Sigma_{\rm SFR}$) of the galaxies of the sample.

The paper is organized as follows. In Sections~\ref{section:sample} and \ref{section:observations} we briefly describe the sample, observations, and data reduction process, which are detailed in Paper I. The procedures for obtaining the emission and \Av\ maps are described in Section~\ref{section:analysis}, and the results and analysis of the \Av\ maps and distributions are presented in Section~\ref{section:results}. Finally, Section~\ref{section:summary} includes a brief summary of the paper. Throughout this work we consider H$_{0}=$70\,km\,s$^{-1}$\,Mpc$^{-1}$, $\Omega_{\rm \Lambda}$ = 0.70, $\Omega_{\rm M}$ = 0.30.

\section{The sample}
\label{section:sample}

\begin{table}[t]
\caption{The SINFONI sample}
\tiny
\centering
{\setlength{\tabcolsep}{1.8pt}
\begin{tabular}{cccccc}
\hline
\hline
     ID1      &   ID2   & z &   D$_{\rm L}$       &     Scale    & log \LTIR \\
Common  &   IRAS  &    &(Mpc)    &(pc/arcsec) & (L$_\odot$) \\
     (1)       &   (2)     &    (3)   &    (4)      &      (5)     &        (6)      \\
\hline
\object{IRASF 12115-4656} & \object{IRAS 12115-4657} & 0.018489 & 84.4 & 394 & 11.10 \\
\object{IC 5179} & \object{IRAS 22132-3705} & 0.011415& 45.6 & 216 & 11.12 \\
\object{NGC 2369} & \object{IRAS 07160-6215} & 0.010807& 48.6 & 230 & 11.17 \\
\object{NGC 5135} & \object{IRAS 13229-2934} & 0.013693& 63.5 & 299 & 11.33 \\
\object{NGC 3110} & \object{IRAS 10015-0614} & 0.016858& 78.4 & 367 & 11.34 \\
\object{NGC 7130} & \object{IRAS 21453-3511} & 0.016151 & 66.3 & 312 & 11.34 \\
\object{ESO 320-G030} & \object{IRAS 11506-3851} & 0.010781& 51.1 & 242 & 11.35\\
\object{IRASF 17138-1017} & \object{IRAS 17138-1017} & 0.017335& 75.3 & 353 & 11.42 \\
\object{IC 4687} & \object{IRAS 18093-5744} & 0.017345 & 75.1 & 352 & 11.44 \\
\object{NGC 3256} & \object{IRAS 10257-4338} & 0.009354 & 44.6 & 212 & 11.74 \\
\object{IRAS 23128-5919} & \object{IRAS 23128-5919}& 0.044601& 195 & 869 & 12.04 \\
\object{IRAS 21130-4446} & \object{IRAS 21130-4446} & 0.092554& 421 & 1712 & 12.22 \\
\object{IRAS 22491-1808} & \object{IRAS 22491-1808} & 0.077760& 347 & 1453 & 12.23 \\
\object{IRAS 06206-6315} & \object{IRAS 06206-6315} & 0.092441& 425 & 1726 & 12.31\\
\object{IRAS 12112+030}5 & \object{IRAS 12112+0305} & 0.073317& 337 & 1416 & 12.38 \\
\object{IRAS 14348-1447} & \object{IRAS 14348-1447} & 0.083000& 382 & 1575 & 12.41 \\
\object{IRAS 17208-0014} & \object{IRAS 17208-0014} & 0.042810& 189 & 844 & 12.43 \\
\hline
\hline
\end{tabular}}
\tablefoot{Col. (3): redshift from the NASA Extragalactic Database (NED). Cols. (4) and (5): Luminosity distance and scale from Ned Wright's Cosmology Calculator \citep{Wright:2006p4236} given h$_{0}$ = 0.70, $\Omega_{\rm M}$ = 0.7, $\Omega_{\rm M}$ = 0.3. Col. (6): \LTIR (8--1000$\mu$m) calculated from the IRAS flux densities $f_{12}$, $f_{25}$, $f_{60}$ and $f_{100}$ \citep{Sanders:2003p1433}, using the expression given in \cite{Sanders:1996p845}.}
\label{table:sample}
\end{table}

The SINFONI sample is a subsample of a larger set of local LIRGs and ULIRGs described in \cite{Arribas:2008p4403} that was observed with different IFS facilities. These facilities include optical and infrared IFS instruments in both hemispheres, such as INTEGRAL+WYFFOS \citep{Arribas:1998p8476} at the 4.2\,m William Herschel Telescope, VLT-VIMOS (\emph{VIsible MultiObject Spectrograph}, \citealt{LeFevre:2003p8480}), PMAS (\emph{Potsdam MultiAperture Spectrophotometer}, \citealt{Roth:2005p4504}), and SINFONI. The large sample ($\sim$70 sources) covers the whole range of LIRGs and ULIRGs IR luminosities and the different morphological classes observed in these objects. 

The sample used for the present study comprises a total of 17 objects, 10 LIRGs, and 7 ULIRGs covering the luminosity range log(\LTIR/\Lsolar)$=11.10-12.43$ (see Table~\ref{table:sample}). It was selected to be representative of the different morphological types of LIRGs and ULIRGs, from isolated galaxies to strongly interacting systems and mergers. The mean redshifts of the LIRG and ULIRG subsamples are $z_{\rm LIRGs}=0.014$ and $z_{\rm ULIRGs}=0.072$ and the mean luminosities are log(\LTIR/\Lsolar)$=11.33$ and log(\LTIR/\Lsolar)$=12.29$, respectively. More details on the sample can be found in Paper I.

\section{Observations and data reduction}
\label{section:observations}
The data were obtained in service mode between April 2006 and July 2008 using SINFONI on the VLT (periods 77B, 78B, and 81B). The sample was observed in the K band (1.95--2.45\,$\mu$m) with a plate scale of 0\farcs125$\times$0\farcs250\,pixel$^{-1}$ that results in an FoV of 8"x8" by a 2D 64x64 spaxel frame. The spectral resolution is R$\sim$4000, and the full-width-at-half-maximum (FWHM) as measured from the OH sky line at 2.190\,$\mu$m is $\sim$6.0\,\AA\ with a dispersion of 2.45\,\AA\ per pixel. The observations have a typical resolution of $\sim$0.63\,arcsec (FWHM, seeing-limited) that corresponds, on average, to $\sim$0.2\,kpc and $\sim$0.9\,kpc for LIRGs and ULIRGs, respectively.

Owing to the limited FoV, the data sample typically $\sim3\times3$\,kpc for the LIRGs and $\sim12\times12$\,kpc for the ULIRGs subsample. Some of the more extended systems were then observed in different pointings to cover regions of interest like secondary nuclei or star-forming complexes. For a detailed description of the observations, pointings, and integration times, see Paper I.

The reduction process was performed using the ESO pipeline ESOREX (version 2.0.5) and our own IDL routines for the flux calibration. All the individual frames were corrected from dark subtraction, flat fielding, detector linearity, geometrical distortion, wavelength calibration and sky-subtraction. After this process, the cubes of those objects with several pointings were combined to build a final mosaic.

The maps of the \Pa, \Brg, and \Brd\ lines were constructed by fitting a Gaussian profile on a spaxel-by-spaxel basis. We have developed our own routines, based on the IDL routine MPFIT \citep{Markwardt:2009p7399}, to perform the fitting of the cubes in an automated fashion. As described in Paper I, the data were binned using the Voronoi method developed by \cite{Cappellari:2003p4908} in order to maximise the S/N over the entire FoV. Each map was binned independently, since the S/N depends on the wavelength, as well as the spatial distribution of the emission, to achieve a minimum S/N on average in the whole FoV. This minimum S/N varies from object to object, and is typically between 15 and 25 for the brightest line (\Brg\ and \Pa\ in LIRGs and ULIRGs, respectively), and between 8 and 10 for the weakest (\Brd\ and \Brg\ for LIRGs and ULIRGs, respectively). Further details on the flux calibration and the individual S/N thresholds used in the Voronoi binning can be found in the Paper I of these series.

\section{Data analysis}
\label{section:analysis}

\begin{figure*}
\begin{center}
\resizebox{0.9\hsize}{!}{\includegraphics[angle=0]{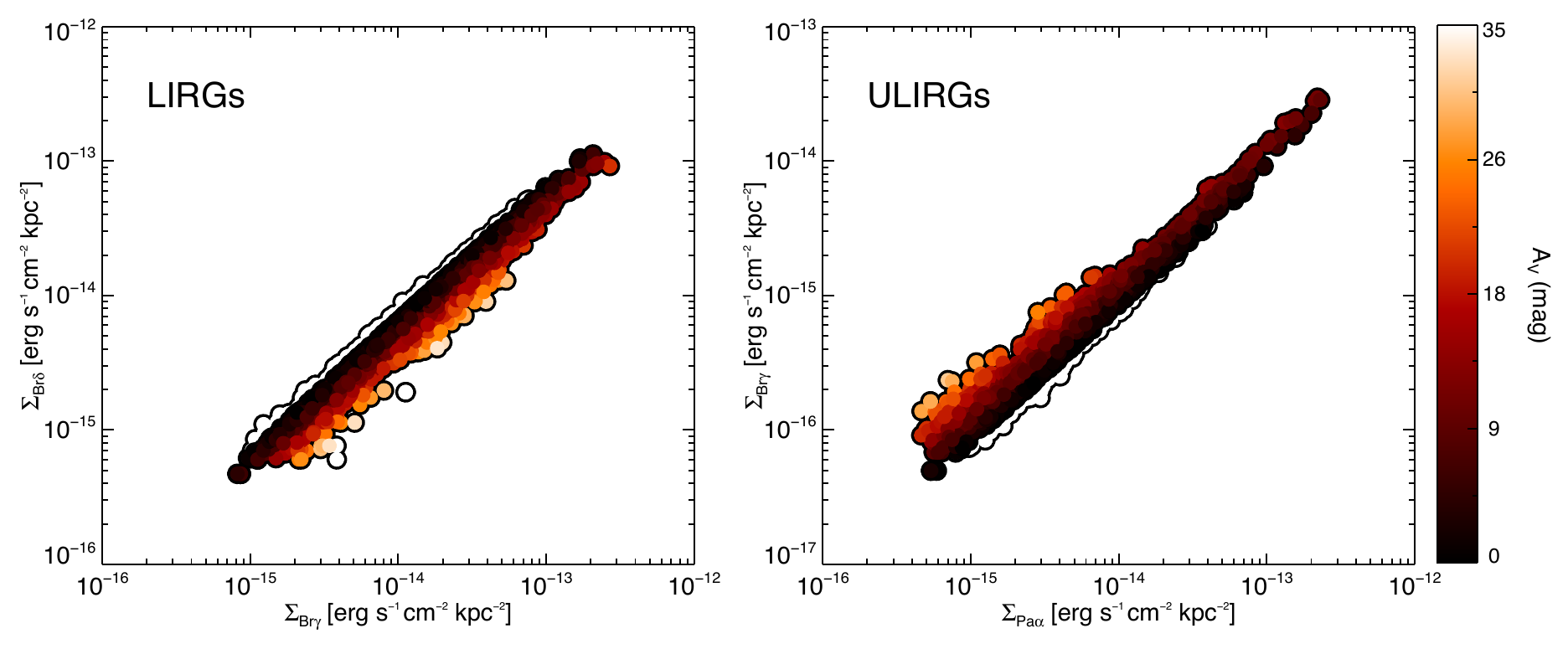}}
\caption{Surface brightness distributions of the individual spaxels of LIRGs and ULIRGs. For clarity, we only plot a random distribution of the 20\% and 50\% of the points for LIRGs and ULIRGs, respectively. The colour code indicates the \Av\ value of each spaxel, whereas the points with \Av$<0$ are outlined with a black contour line. The increase in the extinction towards low surface brightness values is mainly due to the high uncertainties of the flux measurements, in particular in ULIRGs, where the highest \Av\ values correspond to the spaxels with the lower \Brg\ surface brightness.}
\label{figure:s_density}
\end{center}
\end{figure*}

The 2D extinction / dust structure was derived using the \Brg/\Brd\ and \Pa/\Brg\ line ratios for LIRGs and ULIRGs respectively. Although the \Brd\ line is detected in most of the ULIRGs, its S/N is not high enough to map the emission and, in most of the cases, it is not sufficient to perform an integrated analysis of the emission.

As mentioned before, the maps of the different lines were constructed by fitting a single Gaussian profile on a spaxel-by-spaxel basis (see Fig.~\ref{figure:LIRG} and \ref{figure:ULIRG}). Based on the emission maps, we obtained the extinction in magnitudes (\Av) following the procedure outlined in \cite{Bedregal:2009p2426}. We compared the theoretical ratio between the two lines (\Brg/\Brd $=1.52$ and \Pa/\Brg$=12.07$ at T$=10,000$\,K and n$_{\rm e}=10^4\,cm^{-3}$, case B; \citealt{Osterbrock:1989AGN2}) with the measurements for each spaxel. The extinction in magnitudes could be expressed in the form

\begin{equation}
 \rm A_{\rm \lambda_1} - A_{\rm \lambda_2} = -2.5\cdot\rm log\left[\cfrac{\left(F_{\rm \lambda_1}/F_{\rm \lambda_2}\right)_{\rm O}}{\left(F_{\rm \lambda_1}/F_{\rm \lambda_2}\right)_{\rm T}}\right],
 \label{eq1}
\end{equation}

\noindent where F$_{\rm\lambda_{\rm i},O}$ and F$_{\rm\lambda_{\rm i},T}$ are the observed and theoretical fluxes for a line centred at $\rm\lambda_{\rm i}$. We made use of the extinction law described in \cite{Calzetti:2000p2349} to express Equation~\ref{eq1} in terms of the visual extinction \Av\ (A$_{\rm Br\gamma} = 0.096$\,\Av, A$_{\rm Br\delta} = 0.132$\,\Av\ and A$_{\rm Pa\alpha} = 0.145$\,\Av).

Since the individual values of \Av\ are sensitive to the S/N of the weakest line (\Brd\ and \Brg\ for LIRGs and ULIRGs respectively), we have only considered those spaxels where the weakest line has been detected above an S/N threshold of four to obtain reliable \Av. This effect is very significative in the case of the \Brg/\Brd\ ratio since the \Brd\ line lies close to the blue limit of the SINFONI K-band. As discussed in Paper I, this wavelength region is strongly affected by noise due to the sky emission, and the atmospheric transmission also decreases. This translates into a more complex local continuum determination, making the line fitting more uncertain. An excess in the continuum level estimation would decrease the line flux and, therefore, increase the extinction (see expression above). Although the \Pa\ line also lies in this region of the spectra, this effect is not so relevant, given the strength of the line (more than $\times10$ the \Brg\ emission), and it is in the numerator of Equation \ref{eq1}.

The 1$\rm\sigma$ uncertainties of the individual  \Av\ values vary typically from 10--20\% in central regions with high S/N, up to 70--80\% in external areas of low surface brightness ($\Sigma\lsim$10$^{-16}$\,erg\,s$^{-1}$\,cm$^{-2}$\,kpc$^{-2}$), with a median value of 30-35\%. Owing to the larger individual errors in the low surface brightness areas, we observed an artificial increase in the \Av\ measurements, especially in the ULIRGs. This systematic effect is observed in Fig.~\ref{figure:s_density}, where the highest extinction values are measured in those spaxels with the lower \Brg\ surface brightness. The \Av\ uncertainties are obtained from the corresponding errors in the fluxes of the two lines, which as described in Paper I, are estimated using Monte Carlo simulations. The advantage of this kind of error estimation is that uncertainties are directly measured from the spectra, so they not only include the effect from photon noise but also take uncertainties due to an improper continuum determination or line fitting into account.

\begin{table*}
{\small
\caption{Integrated properties and statistics of the \Av\ distributions}
\centering
{\setlength{\tabcolsep}{3pt}
\begin{tabular}{cx{0.6cm}@{ $\pm$ }z{0.6cm}x{0.6cm}@{ $\pm$ }z{0.6cm}x{0.6cm}@{ $\pm$ }z{0.6cm}x{0.6cm}@{ $\pm$ }z{0.6cm}cx{0.6cm}@{ $\pm$ }z{0.6cm}ccc}
\hline
\hline
Object  & \multicolumn{2}{c}{A$_{\rm V,nuclear}$} & \multicolumn{2}{c}{A$_{\rm V,nuclear}$} & \multicolumn{2}{c}{A$_{\rm V,R_{\rm eff}}$}& \multicolumn{2}{c}{A$_{\rm V,FoV}$} & A$_{\rm V,median}$ & \multicolumn{2}{c}{A$_{\rm V,mean}$} & A$_{\rm V}$\,(P$_{5}$)& A$_{\rm V}$\,(P$_{95}$) & N$_{\rm spaxel}$ \\
 (1) & \multicolumn{2}{c}{(2)}& \multicolumn{2}{c}{(3)} & \multicolumn{2}{c}{(5)}& \multicolumn{2}{c}{(6)} & (7) &\multicolumn{2}{c}{(8)}  & (9) & (10) & (11)\\ 
\hline
    IRASF12115-4656 &    \multicolumn{2}{c}{\nodata} &    \multicolumn{2}{c}{\nodata} &    7.7 &    0.1 &   5.2 &    0.1 &    4.9 &   6.3 &    8.1 &  -3.65 &   21.5 &   1533 \\
              IC5179 &   10.4 &    0.4 &    \multicolumn{2}{c}{\nodata} &    4.2 &    0.1$^{\ddag}$ &   4.2 &    0.1 &    4.4 &   5.2 &    6.3 &  -3.00 &   17.3 &   2099 \\
             NGC2369 &   16.8 &    0.2 &    \multicolumn{2}{c}{\nodata} &   17.3 &    0.1 &  15.0 &    0.1 &   15.5 &  13.5 &    6.6 &   3.75 &   26.0 &    603 \\
             NGC5135 &    2.8 &    0.2 &    \multicolumn{2}{c}{\nodata} &    3.8 &    0.1 &   3.7 &    0.1 &    4.2 &   4.7 &    4.5 &  -1.72 &   11.7 &   1100 \\
             NGC3110 &    \multicolumn{2}{c}{\nodata} &    \multicolumn{2}{c}{\nodata} &    7.4 &    0.1$^{\ddag}$ &   7.4 &    0.1 &    7.9 &   7.9 &    5.8 &  -3.02 &   16.6 &    400 \\
             NGC7130 &   12.3 &    0.3 &    \multicolumn{2}{c}{\nodata} &   11.8 &    0.2 &   8.7 &    0.1 &    6.2 &   5.9 &    6.5 &  -2.45 &   18.6 &    687 \\
         ESO320-G030 &    \multicolumn{2}{c}{\nodata} &    \multicolumn{2}{c}{\nodata} &    7.8 &    0.1 &   7.0 &    0.1 &    7.8 &   8.0 &    7.1 &  -2.82 &   20.1 &   1383 \\
     IRASF17138-1017 &   15.2 &    0.4 &    \multicolumn{2}{c}{\nodata} &   11.1 &    0.1 &   7.0 &    0.1 &    7.6 &   6.1 &    5.6 &  -0.91 &   17.2 &    806 \\
              IC4687 &   10.6 &    0.1 &    \multicolumn{2}{c}{\nodata} &    3.9 &    0.1 &   3.7 &    0.1 &    4.2 &   4.5 &    5.1 &  -1.74 &   13.1 &   3401 \\
    NGC3256$^{\dag}$ &    \multicolumn{2}{c}{\nodata} &   12.2 &    0.2 &    5.0 &    0.1 &   4.1 &    0.1 &    5.0 &   6.6 &    6.1 &  -2.73 &   16.8 &   4522 \\
      IRAS23128-5919 &    8.7 &    0.3 &    6.7 &    0.2 &    7.1 &    0.1 &   6.8 &    0.1 &    6.2 &   7.0 &    4.7 &  -1.60 &   13.4 &   1182 \\
      IRAS21130-4446 &    4.3 &    0.4 &    3.2 &    0.2 &    4.7 &    0.3 &   4.4 &    0.1 &    5.4 &   5.0 &    5.9 &  -1.80 &   15.5 &    214 \\
      IRAS22491-1808 &    4.1 &    0.3 &    4.5 &    0.2 &    5.5 &    0.1 &   5.3 &    0.1 &    6.4 &   6.5 &    5.1 &  -1.79 &   16.0 &    299 \\
      IRAS06206-6315 &    7.5 &    0.3 &   11.3 &    0.4 &    7.6 &    0.2 &   8.5 &    0.2 &   10.7 &   9.5 &    7.7 &  -3.05 &   22.7 &    114 \\
      IRAS12112+0305 &    8.9 &    0.3 &    8.4 &    0.2 &    9.2 &    0.2 &   8.5 &    0.1 &    8.2 &   9.0 &    6.2 &  -0.52 &   20.9 &    466 \\
      IRAS14348-1447 &    5.5 &    0.2 &    8.1 &    0.3 &    6.2 &    0.2 &   7.1 &    0.1 &    8.6 &   7.9 &    6.9 &  -1.04 &   20.6 &    301 \\
      IRAS17208-0014 &    8.0 &    0.2 &    \multicolumn{2}{c}{\nodata} &    6.8 &    0.1 &   5.8 &    0.1 &    4.8 &   6.3 &    4.0 &  -1.61 &   10.8 &    485 \\
\hline
\hline
\end{tabular}}
\tablefoot{Cols. (2) and (3):  Nuclear extinction measured at the main (2) and secondary (3) nucleus within an aperture radius of 0$\farcs$63. Cols. (4) and (5): Measurements of the \Av\  within the H$\alpha$ effective radius from \cite{Arribas:2012p1203}, and the integrated measurement over the whole FoV, respectively. Cols. (6) to (9): Median, weighted mean, and 5th and 95th percentiles of the \Av\ distributions. Col. (10): Number of valid spaxels in the \Av\ maps and distributions.
$^{\dag}$ Since the main nucleus of \object{NGC 3256} was not observed, we centred the aperture for A$_{\rm V,R_{\rm eff}}$ on the centre of the FoV, so the measurement might be inaccurate.
$^{\ddag}$Due to the limited FoV of the observations, in these objects the H$\alpha$ effective radius is greater than our FoV, and the A$_{\rm V,R_{\rm eff}}$ measurements are equivalent to A$_{\rm V,FoV}$.}
\label{table:av_table}
}
\end{table*}

\section{Results and discussion}
\label{section:results}
\subsection{Two-dimensional extinction structure in LIRGs and ULIRGs}
\label{section:structure}

\begin{figure}[t]
\begin{center}
\resizebox{\hsize}{!}{\includegraphics[angle=0]{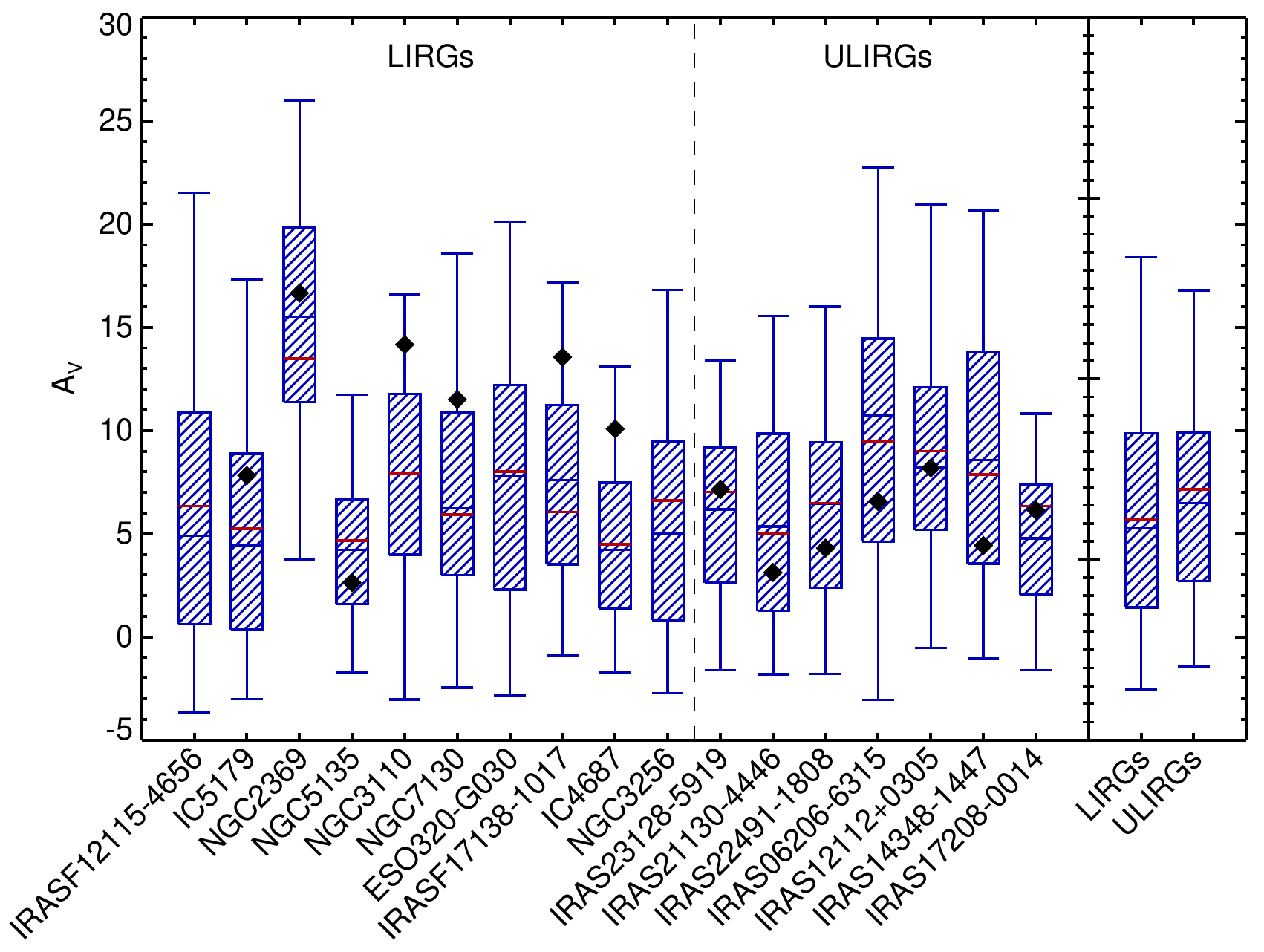}}
\caption{Individual \Av\ distributions, on a spaxel-by-spaxel basis, of the galaxies of the sample ordered by increasing \LTIR. The extremes of the distributions are the 5th and 95th percentiles (P$_5$ and P$_{95}$). The boxes illustrate the interquartile range, whereas the horizontal blue and red lines correspond to the median and the weighted mean of the distribution, respectively. The measurements of the nuclear extinction are plotted as black diamonds. The total \Av\ distributions for LIRGs and ULIRGs are shown on the right-hand side of the plot.}
\label{figure:av_distributions}
\end{center}
\end{figure}

\begin{figure*}[t]
\begin{center}
\resizebox{\hsize}{!}{\includegraphics[angle=0]{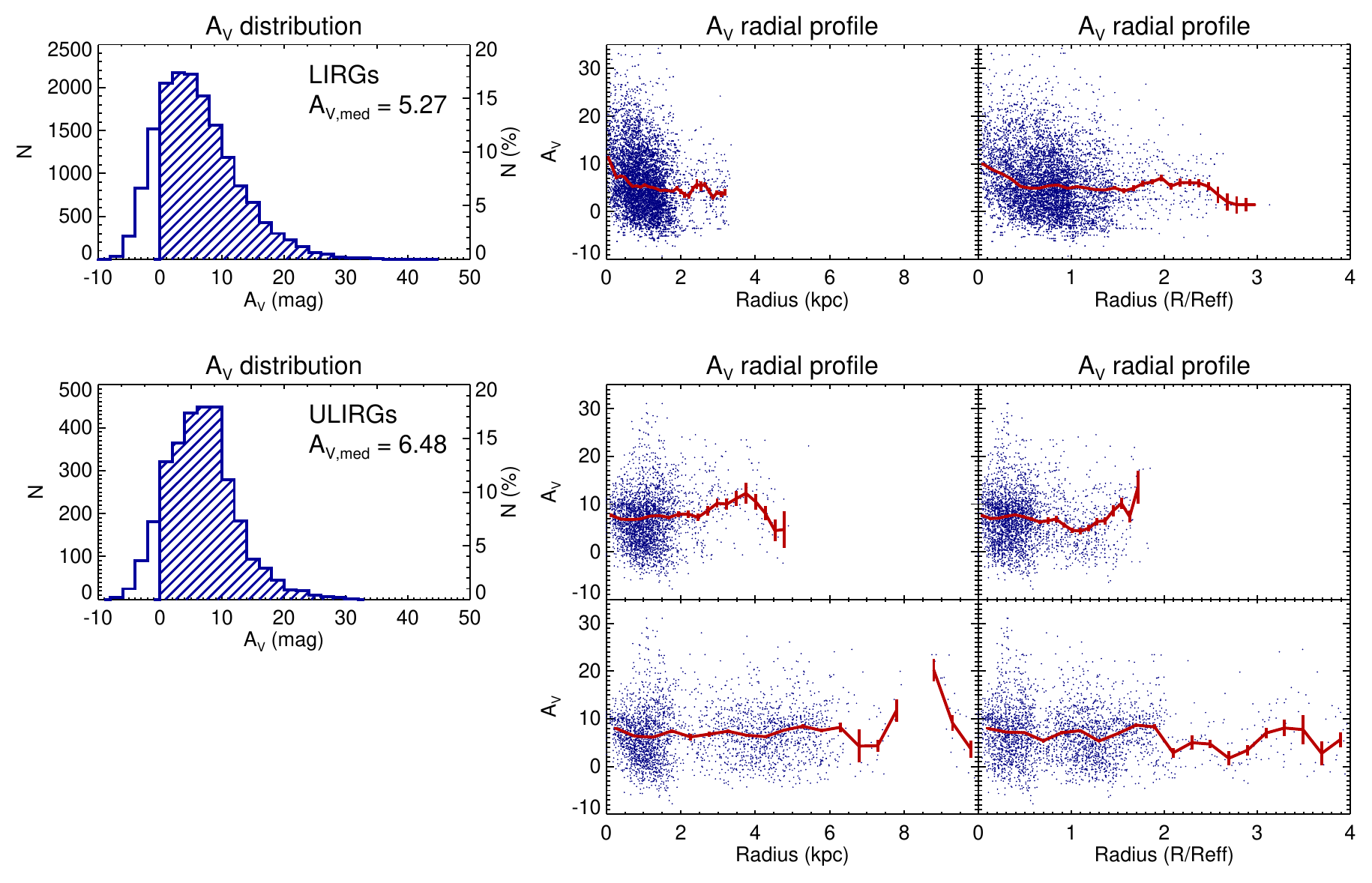}}
\caption{\Av\ distributions and radial profiles of the LIRGs (top) and ULIRGs (bottom) subsamples on a spaxel-by-spaxel basis. The median \Av\ values of each distribution are shown in the left panels. In the central and right panels, the radial profiles are plotted in terms of the radius in kpc (centre) and in units of the H$\alpha$ effective radius (R$_{\rm eff}$, right panel) extracted from \cite{Arribas:2012p1203}. The red line represents the weighted mean of \Av\ and its error for different radial bins in steps of 1/30 of the total radial coverage. For the ULIRG subsample, we plotted the radial profile of the sample by considering each component of the systems separately (top) and the profile extracted by taking the brightest nucleus in the K-band continuum as the centre of the systems (bottom).}
\label{figure:samples_distributions}
\end{center}
\end{figure*}

Our 2D extinction maps cover areas from $\sim$2\,kpc to $\sim$12\,kpc for LIRGs and ULIRGs, respectively (see Figs.~\ref{figure:LIRG} and \ref{figure:ULIRG}). The seeing-limited observations provide a linear resolution equivalent to a physical scale resolution of $\sim$0.2\,kpc and $\sim$0.9\,kpc for each subsample. On these scales, the extinction maps show that dust is not uniformly distributed, revealing a clumpy structure with almost transparent areas (\Av$\lsim$1\,mag) and regions where the visual extinction is higher than ten magnitudes.

In LIRGs, the extinction maps show a very irregular and clumpy structure on scales of $\sim$200-300\,pc, already observed from near-IR continuum maps \citep{Scoville:2000AJ119}. The higher \Av\ values are usually associated with the nuclear regions of the objects, although obscured extranuclear regions are also common. The star-forming regions with high \Brg\ surface brightness, found along the dynamical structures like arms of rings, are typically low-extinction regions.

As shown in the radial profiles of \object{NGC 3110} or \object{IRASF 12115-4656} (Figs.~\ref{figure:NGC3110} and \ref{figure:IRASF12115}), the extinction increases inwards up to \Av$\sim$15--25\,mag. This behaviour is also observed in \object{ESO 320-G030} (Fig.~\ref{figure:ESO320}), where the emission is concentrated on a star-forming ring of $\sim$500--600\,pc radius. Although the nucleus might also be obscured, the radial profile of this object is not as steep as in \object{NGC 3110} or \object{IRASF 12115-4656} and indicates that the lack of emission could also be due to the intrinsic distribution of the star-forming regions around the ring.

The morphology of the \Av\ maps in the ULIRG subsample suggest a patchy, non-uniform distribution of the dust, typically on physical scales of $\gsim$1\,kpc that correspond to our resolution limit. Owing to the higher linear resolution (i.e. kpc/spaxel) of the ULIRG subsample, the comparison with the LIRGs is not straightforward. As shown in Fig.~\ref{figure:samples_distributions}, although our data samples similar areas of $\sim$1-2\,\Reff, the dust structure is probed with significantly different spatial resolutions, owing to the factor $\times$5 in distance between both subsamples. This difference precludes our from resolving sub-kiloparsec structures in ULIRGs, such as the ones observed in the LIRG \Av\ maps, limiting our physical resolution to $\sim$1\,kpc. As discussed in Sec.~\ref{section:scale}, these differences in the linear resolution between both subsamples not only shape the observed dust morphology in the more distant galaxies, but also might determine global measurements of the extinction, such as the median of the \Av\ distributions. In Sec.~\ref{section:high-z} we discuss how this distance effect would have direct implications for the study of high-z galaxies.

\subsection{A$_V$ distributions and radial profiles}
\label{section:distrib}

Figures~\ref{figure:LIRG} and \ref{figure:ULIRG} show the \Av\ distributions for each galaxy. Although it is clear that most of the spaxels with \Av$<$0 have no physical meaning individually, we have kept them in the distributions since they do have statistical relevance. If we remove them from the distributions, we introduce a bias toward the high \Av\ values, displacing the mean and median of the distributions artificially. On the other hand, due to the S/N threshold adopted, we assure that most of those spaxels with \Av$<$0 are compatible with \Av$\sim$0 within the uncertainties.

The histograms show a wide variety of distributions, from narrow, peaked distributions, such as \object{NGC 5135} or \object{IC 4687} (Fig.~\ref{figure:NGC5135} and \ref{figure:IC4687}), concentrated towards low \Av\ values, to wide distributions such as \object{NGC 3110} or \object{IRAS 14348-1447} (Figs.~\ref{figure:NGC3110} and \ref{figure:IRAS14348}) that extend up to $\sim$30-35\,mag. The median and weighted mean \Av\ values, together with the 5th and 95th percentiles of the distributions, are listed in Table~\ref{table:av_table}. Figure \ref{figure:av_distributions} also compares the individual distributions of each galaxy of the sample, ordered by increasing  \LTIR. As shown in the figure, most of the individual values, within the interquartile ranges, are concentrated between \Av$\sim$1 and \Av$\sim$20 mag, and there is no clear evidence of any dependence with \LTIR.

In LIRGs, the visual extinction ranges between \Av$\sim1-20$\,mag, whereas in ULIRGs, the \Av\ values range between \Av$\sim2-15$\,mag. In LIRGs, these values of the visual extinction are very similar although slightly lower than previous results in the mid-infrared from Spitzer, \Av$\lsim1-30$\,mag with a mean value of $\sim$11\,mag (\citealt{PereiraSantaella:2010kj}, \citealt{Alonso-Herrero:2012p744}). On the other hand, ULIRGs show slightly higher values than previous measurements in the optical, from \Av$\lsim$0.2\,mag to $\sim9$\,mag \citep{GarciaMarin:2009p8459}, and significantly lower than mid-infrared measurements based on the silicate absorption feature at 9.7\,$\mu$m, from \Av$\sim$6\,mag up to \Av$\gsim$40\,mag \citep{Imanishi:2007p2639}. 

These \Av\ values have to be considered as lower limits of the dust extinction, since the theoretical values of the ratios are based on the assumption that the gas is optically thin.  However, it is well known that the central regions of these objects are dusty environments and that the observed recombination lines might have been originated at different depths within the star-forming regions. That also means that we would infer different extinction values from different emission line ratios, since the lines are probing different optical depths.

To compare the \Av\ distributions of LIRGs and ULIRGs, we combined all the available spaxels of each luminosity bin to obtain a typical \Av\ distribution. These distributions are shown in Fig.~\ref{figure:av_distributions} and, in more detail, in Fig.~\ref{figure:samples_distributions}. The median values of \Av\ for each subsample are similar, \Av$_{\rm med}=5.3$\,mag and \Av$_{\rm med}=6.5$\,mag, and the number of individual spaxels is $\sim$16000 for the LIRG distribution and $\sim$3000 for the ULIRG one. The shape of the distributions are, however, slightly different. Whereas the LIRG distribution seems to have more than 50\% of the points concentrated within the range \Av$\simeq1-10$\,mag, the ULIRGs extend over a somewhat narrower range of \Av$\simeq3-10$\,mag and tend to reach higher \Av\ values on individual spaxels. On the other hand, the modes of the distributions are also different; whereas the LIRG distribution peaks at $\sim3-4$\,mag, the mode of the ULIRG distribution reaches up to $\sim7-8$\,mag. Although some of these differences might be intrinsic, we discuss in Section~\ref{section:scale} that the physical sampling of the maps plays an important role in the study of the extinction, owing to the patchy structure of the dust.

We obtained the radial profiles of the extinction for every individual object of the sample and characteristic profiles for both LIRG and ULIRG subsamples. For this purpose, we adopted the same criterion as in Paper I to identify the central spaxel of each object with the brightest spaxel of the FoV in the K-band image (see Figs.\ref{figure:LIRG} and \ref{figure:ULIRG}). The only exception in \object{NGC 3256}, since the nucleus of this galaxy was not observed in the K-band (see Paper I for further details). For this object, we used the H-band continuum image of the nucleus to identify the central spaxel. For those systems with multiple nuclei (i.e. all the ULIRGs with the exception of \object{IRAS 17208-0014}), we also extracted the radial profiles of each component separately.

As shown in Figs.~\ref{figure:LIRG} and \ref{figure:ULIRG}, the profiles typically sample the inner $\sim2$\,kpc for the LIRGs and $\lsim$6\,kpc for the ULIRGs, and most of them have an almost flat or negative slope. LIRGs show steeper negative slopes than ULIRGs, especially in the central 0.5--1\,kpc. In LIRGs, these slopes are typically of $\sim-$2.4\,mag\,kpc$^{-1}$ on average, versus $\sim-$0.3\,mag\,kpc$^{-1}$ in ULIRGs. This could be explained by the different sampling scales of both subsamples, since we cannot resolve the innermost regions of the ULIRGs with the resolution achieved for the LIRG subset. In those pre-coalescent systems with multiple nuclei, we found no systematic differences between the radial profiles of both components. These profiles typically sample the innermost $\sim2-3$\,kpc of each component of the system, and show no clear radial dependence of the extinction on these scales. 

We extracted the radial profiles of the total \Av\ distributions of LIRGs and ULIRGs separately. Since each object has been observed with different sampling and physical scales, we also obtained the profiles in units of the effective radius \Reff, using the values from \cite{Arribas:2012p1203} obtained from H$\alpha$ maps. As shown in Fig.~\ref{figure:samples_distributions}, the LIRG profile is almost flat beyond r$\sim$1\,kpc or $\sim$0.5\,\Reff, with an average value of \Av$\sim$5.3\, mag. Within the central kiloparsec, the extinction increases up to $\sim$10\,mag. The ULIRG subsample shows a very uniform profile, with a median value of \Av$\sim$6.0\,mag and only local deviations due to the presence of double nuclei in some of the systems, or due to strong complexes of star formation at distances beyond 2-3\,kpc radius (see Fig.\ref{figure:IRAS06206} and \ref{figure:IRAS12112} for some examples). The radial profile shows small differences when extracted for each component separately, with an almost flat slope over the inner 2-3\,kpc radius ($\sim$\Reff). The measurements beyond $r\sim$6\,kpc or $r\sim$2\,\Reff\ are very uncertain, owing to the lack of available spaxels and to the low surface brightness of the \Brg\ emission in these regions. As mentioned before, the extinction in ULIRGs derived using Eq.~\ref{eq1} is highly affected by noise fluctuations of the \Brg\ line.

\subsection{Nuclear and integrated A$_V$ measurements}
\label{section:nuclear}

We obtained integrated \Av\ measurements for different regions of interest in each object (Table~\ref{table:av_table}). The uncertainties of the parameters are obtained by a Monte Carlo method of $\rm N=1000$ simulations, and do not take the 1$\sigma$ uncertainties of $\sim$5\% into account in the absolute flux calibration (see \citealt{Piqueras2012A&A546A}). We found that, in $\sim$70\% of the objects, its nucleus corresponds to a peak in the extinction, ranging from \Av$\sim$3\,mag up to \Av$\sim$17\,mag. These values of the nuclear extinction are higher than each median and mean \Av\ values in $\sim57$\% of the objects. However, there are some galaxies, such as \object{NGC 3110} or \object{IC 4687} (Figs.~\ref{figure:NGC3110} and \ref{figure:IC4687}), where the nucleus is completely obscured, and no measurements of the extinction are available. That the nucleus of the objects coincides with a peak of the extinction agrees with other studies in LIRGs and ULIRGs (see \citealt{AlonsoHerrero:2006p4703} and \citealt{GarciaMarin:2009p8459}) based on H$\alpha$/H$\beta$ and Pa$\alpha$/\Brg\ ratios, and with mid-infrared studies with Spitzer, based on the silicate absorption feature at 9.7\,$\mu$m (\citealt{Imanishi:2007p2639}, \citealt{PereiraSantaella:2010kj}). These authors found that either the highest extinctions coincide with the nucleus of the objects or the nuclear regions are local maxima in the \Av\ maps.

For those objects with a double nucleus, we also presented integrated measurements of the extinction for the secondary one. Although almost all of the interacting systems in our sample are ULIRGs, \object{NGC 3256} presents a well known, highly obscured nucleus $\sim5$\,arcsec to the south of the main one. Our measurements of the extinction of the southern nucleus of this object shows that it is one of the most extinguished regions in our sample, with \Av$\simeq$12.2\,mag, in good agreement with previous works (\citealt{Kotilainen:1996p305}, \citealt{AlonsoHerrero:2006p4703}, \citealt{DiazSantos:2008p685}, \citealt{Rich:2011ApJ734}).

\subsection{Dust clumpiness and the effect of the linear resolution.}
\label{section:scale}

\begin{figure}
\begin{center}
\resizebox{\hsize}{!}{\includegraphics[angle=0]{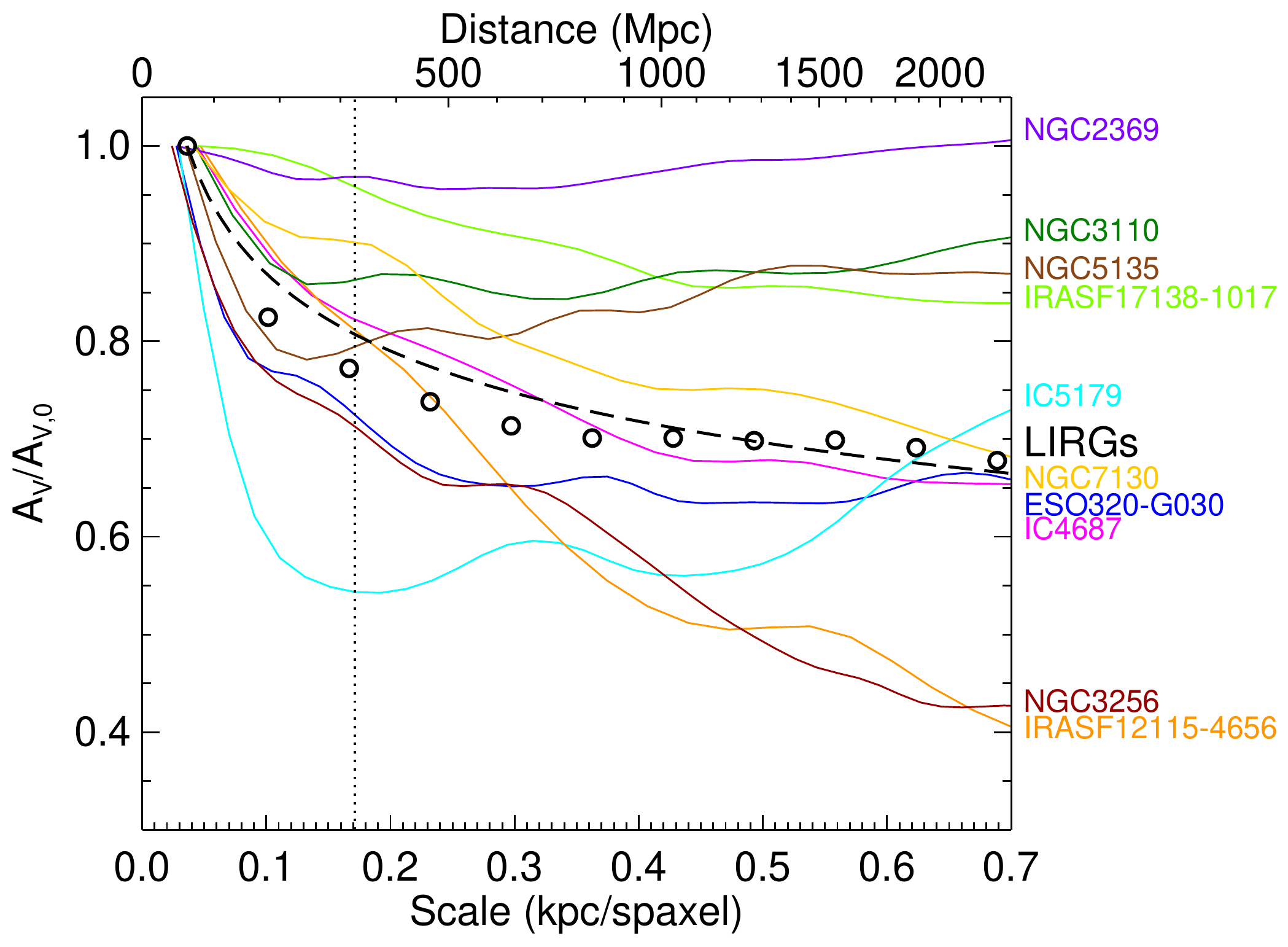}}
\caption{Evolution of the computed median \Av\ of the individual distributions of the LIRG subsample as a function of scale/distance. The \Av\ values are normalised to the rest-framed distribution. The value that corresponds to the \Av\ distribution of the whole LIRG subsample is plotted as black circles and a power-law fit to the data as a dashed black line. The mean distance of the ULIRG subsample (328\,Mpc) is plotted as a vertical dotted line for reference.}
\label{figure:sim_individual}
\end{center}
\end{figure}

\begin{figure*}[t]
\begin{center}
\resizebox{\hsize}{!}{\includegraphics[angle=0]{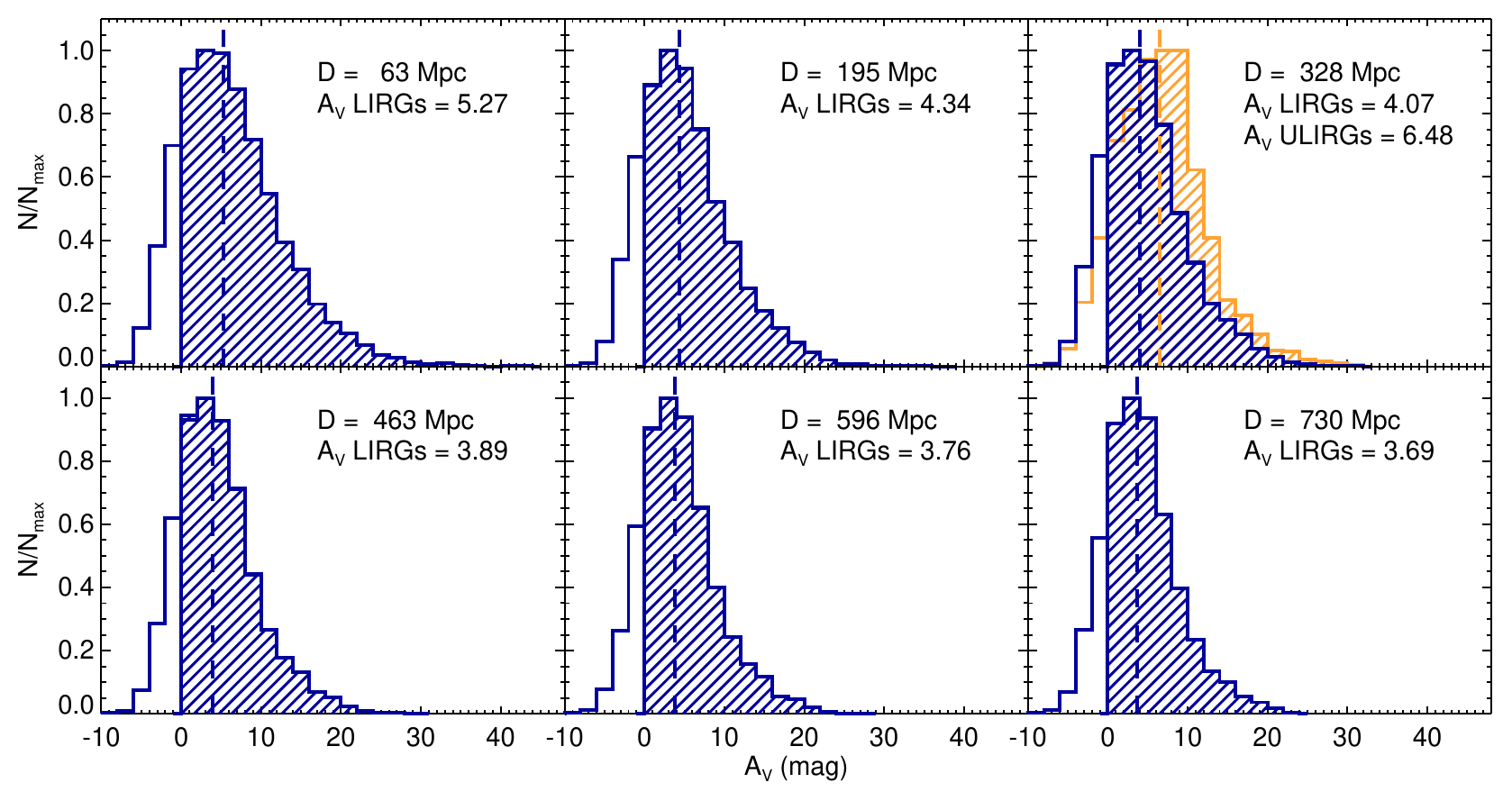}}
\caption{Different simulations of the LIRGs \Av\ distribution for increasing distance (from top to bottom and left to right). The median of each distribution is highlighted with a vertical dashed line. The distance and the median \Av\ of the simulated distribution is annotated in each panel. The top left panel corresponds to the average distance of the ULIRG subsample, and shows the observed \Av\ distribution of the ULIRGs and its median, in yellow.}
\label{figure:sim_distributions}
\end{center}
\end{figure*}

As mentioned in Section~\ref{section:structure}, the \Av\ maps reveal a patchy, clumpy structure of the dust at sub-kiloparsec and kiloparsec scales in LIRGs and ULIRGs, respectively. Given this non-uniform distribution, the measurements of the extinction at different distances might be affected by the physical scale of the observations. To probe how the pixel scale might bias the \Av\ measurements, we have obtained the \Av\ distribution of our LIRG sample simulating different scales, hence different distances. This distance effect because of the linear resolution might be even more relevant for high-z objects, where the structure is sampled on even larger scales of $\sim$1-2\,kpc.

To simulate the distribution at further distances, we degraded the individual \Brg\ and \Brd\ maps to different scales, and obtained maps of poorer spatial resolution. In this process, we only considered the same valid spaxels as in the original maps. Once the maps were degraded, we obtained the \Av\ distributions of each individual object as described in Section \ref{section:analysis}, which were finally combined in one single distribution.

Since the FoV of our SINFONI observation is limited to $\sim8"\times8"$, we could only sample the innermost $\sim3\times3$\,kpc of the LIRGs (typically $\sim$R$_{\rm eff}$). If we translate these scales to a distance that is ten times larger, these $\sim3$\,kpc are equivalent to an angular distance of $\sim$1", i.e. 1/8 of the FoV or $\sim8$\, spaxels. The lack of data from the external regions of the local objects keeps the extrapolation to larger distances from beeing straight-forward.

Figure~\ref{figure:sim_individual} shows the relation of the median of the individual \Av\ simulated distributions to the sampling scale/distance, for each galaxy in the LIRG subsample. Although each curve shows a different behaviour that depends on the particular gas and dust distribution, there is a general trend towards lower \Av\ values as distance increases. This decrement of the median value of the visual extinction is also observed when we consider the distribution of LIRGs as a class, adding all the individual spaxels of the LIRG subsample in a single distribution, as shown in the figure. To parametrize this observed decrement of the median \Av\ of the LIRG distribution with the sampling scale/distance, we fitted the data points to a simple power-law model, and found that the best-fit model corresponds to

\begin{equation}
\rm A_{\rm V} / A_{\rm V0} \simeq \rm (D/D_{\rm 0})^{-0.13} \simeq \rm (S/S_{\rm 0})^{-0.14},
 \label{eq2}
\end{equation}

\noindent where A$_{\rm V}$ and S are the median values of the \Av\ distribution and the physical scale at a distance D, respectively, and A$_{\rm V0}=5.27$\,mag and S$_{\rm 0}=290$\,pc\,arcsec$^{-1}$ are the median values of the rest-frame \Av\ distribution and median physical scale at the mean distance of the LIRGs subset, D$_{\rm 0}=63.3$\,Mpc, respectively.

In figure~\ref{figure:sim_distributions}, we show in detail the observed \Av\ distribution of the LIRG subsample and the simulated distributions for increasing distances. The different panels reveal that not only the median value of the distribution decreases when the galaxies are sampled on a larger physical scale, but also that the shape of the distribution is slightly different, becoming narrower than the rest-framed distribution, and more compact. The mode of the simulated distribution also changes with respect to the ULIRG observed distribution, and becomes lower than the median. As shown in Fig.~\ref{figure:sim_individual}, the difference between the rest-frame extinction, A$_{\rm V0}$, and the simulated \Av\ increases rapidly within the first D$\sim$700\,Mpc, and seems to reach an asymptotic value of \Av/A$_{\rm V0}\sim$0.65 beyond that distance.

The non-uniform distribution of the dust, even on small scales, makes that, at a given resolution unit, we map both obscured regions and areas where the interstellar medium is more transparent. This average is biased towards the brightest regions, which are those were the emission is less absorbed. The result is that, on average, the dust distribution is smoothed, becomes narrower, and the \Av\ values that we measure are biased towards the lowest values.

As mentioned in Section~\ref{section:distrib}, we measured a median extinction of \Av$_{\rm, LIRGs}=5.3$ and \Av$_{\rm, ULIRGs}=6.5$ magnitudes for our LIRG and ULIRG subsamples, respectively. Since the average distance of each subsample is $\sim63$\,Mpc and $\sim328$\,Mpc (see Paper I), this factor $\sim$5 in distance and sampling scale (from $\sim$0.2\,kpc to $\sim$0.9\,kpc on average, respectively) translates to a decrease in the measured \Av\ of $\sim$1.2\,mag, as shown in Fig.~\ref{figure:sim_distributions} (top right panel). If we assume that LIRGs and ULIRGs have similar structures and apply the same correction to the ULIRGs subsample, we find that the average \Av\ in the ULIRGs, corrected from distance effects, reaches \Av$\sim$8.0\,mag, i.e. $\sim$2.8\,mag higher than in LIRGs.

Besides this resolution/distance effect, the difference between the observed \Av\ distribution in LIRGs and ULIRGs could also be interpreted in terms of intrinsic differences in the morphology between both classes. Although there seems to be a general trend for LIRGs as a class that suggest that the observed \Av\ decreases with the distance/resolution, Fig.\ref{figure:sim_individual} also shows that the behaviour of each individual galaxy is very different, and depends strongly on the particular morphology of the dust distribution. These differences among the objects of the LIRGs subset suggest that the correction to the observed \Av\ could not be accurate for individual galaxies, and would only reflect a general trend for LIRGs as a class.

\subsection{Implication for extinction-corrected properties in high-z galaxies}
\label{section:high-z}

There are different observational proofs that suggest that sub-millimetre galaxies (SMGs) could be the high-z analogous of local (U)LIRG. The sub-millimetre and radio fluxes of this population of galaxies indicate that their bolometric luminosities are comparable to local ULIRGs, whereas their mid-IR emission seems to be similar to the observed in local LIRGs  (\citealt{Kovacs:2006ApJ650}, \citealt{Takata:2006ApJ651}, \citealt{MenendezDelmestre:2009ApJ699}). Besides this, recent IFS-based observations of SMGs reveal that they present evidence of clumpy star formation on kiloparsec scales and similar star-formation rate surface densities (\Si) to the local counterparts (\citealt{Nesvadba:2007ApJ657}, \citealt{Harrison:2012MNRAS426}, \citealt{Alaghband-Zadeh:2012MNRAS424}, \citealt{MenendezDelmestre:2013ve}). It is well known that these high-z galaxies may be highly obscured and that a combination of intense dust-obscured star formation and dust-enshrouded AGN activity would be responsible for the high infrared luminosities of these objects \citep{Blain:2002PhR369}. HST-NICMOS and ACS observations reflect structured dust obscuration in these objects \citep{Swinbank:2010MNRAS405}. In particular, \cite{Takata:2006ApJ651} find that the internal extinction in these objects are similar to the extinction in local ULIRGs, and measured a median extinction of \Av$=2.9\pm0.5$\,mag in a sample of SMGs at z$\sim$1.0--3.5 using the H$_{\alpha}$/H$_{\beta}$ flux ratio. 

Clumpy and dusty star-forming structures have also been identified at high redshifts in more common star forming galaxies (i.e. the so called Main Sequence star forming galaxies). These galaxies have (sub)kpc star-forming clumps mostly spread over galactocentric distances of few to several kpc (\citealt{ForsterSchreiber:2011p731}, \citeyear{ForsterSchreiber:2011p739}, \citealt{Genzel:2011ApJ733}), and with internal nebular extinctions of 2 to 4 magnitudes, assuming $\rm A_{v,H_{\alpha}} = A_{v, stellar}/0.44$ (\citealt{ForsterSchreiber:2009p706}, \citealt{Wuyts:2011ApJ738}).

In the previous subsection we showed that the distance/linear scale may play an important role in deriving the internal extinction properties of LIRGs and ULIRGs, by comparing both populations of galaxies locally, albeit with a difference of a factor $\times$5 in distance between both subsamples. According to the simulations, the median extinction measured in a given galaxy decreases when placed at  increasing distances. This effect is due  to the fixed angular resolution of the IFS data  that translates into a larger physical scale per spatial resolution element (spaxel) as the distance to the galaxy increases, and is particularly important when the physical scales sampled by the spaxel are much larger than that of the intrinsic clumpy structure of  the dust distribution and star-forming regions.

It is clear that if the intrinsic star-forming structure of high-z galaxies is in general similar to that of our local LIRGs (i.e. mostly disks) and ULIRGs (i.e. mostly interacting), the smearing effect mentioned above would have a direct impact in the derivation of their 2D internal extinction values, and of  all relevant subsequent extinction corrected properties such as star formation surface densities, KS-laws, and overall star formation rates. This would certainly be the case not only with seeing-limited near-IR IFS, heavily undersampling  the galaxies at redshifts of 1 to 3, with each spaxel corresponding to about 1.5-2 kpc, but even when using AO assisted IFS where  the spaxel (50 to 100 mas)  translates to about 0.4 to 0.8 kpc. Thus, if the results given by  Eq.~\ref{eq2} (see also Fig.~\ref{figure:sim_individual}) are applied to  high-z galaxies, the extinction values derived directly from the observed optical emission line ratios would require to be increased by a factor 1.4 on average. This correction would correspond to an additional increase in the H$_{\alpha}$ flux by a factor $\sim$3 and hence, in the extinction-corrected SFR and \Si.

Finally, it is worth noticing that this distance effect could also have a minor wavelength dependency. The method presented here to derive the \Av\ values is based on specific emission line ratios in the near-infrared. Since these lines originate in regions of higher optical depths than the optical emission lines, and the stellar continuum measured at optical wavelengths, differences in the amount and/or distance dependency could appear. It would be worth exploring these effects with a suitable set of data, in particular if, as in many high-z studies, the \Av\ corrections applied to the observed H$\alpha$ flux is obtained indirectly using the standard Calzetti recipe A$_{\rm H\alpha}$ = 7.4\,E(B--V), where E(B--V)$_{gas}$ = 0.44\,E(B--V)$_{stars}$ (\citealt{Calzetti:2000p2349}, \citeyear{Calzetti:2001PASP113})

\section{Summary}
\label{section:summary}

In this paper, we presented a detailed 2D study of the extinction structure of a representative local sample of 10 LIRGs and 7 ULIRGs, based on VLT-SINFONI IFS K-band observations. We sample the central $\sim3\times3$\,kpc for LIRGs and the $\sim12\times12$\,kpc for ULIRGs, with average linear resolutions (FWHM) of $\sim$0.2\,kpc and $\sim$0.9\,kpc, respectively. The extinction maps are based on measurements of the \Brg/\Brd\ line ratio for LIRGs and of the \Pa/\Brg\ line ratio for ULIRGs.

In agreement with previous studies, we found that the distribution of the dust in these galaxies presents a patchy structure on sub-kiloparsec scales, with regions almost transparent with \Av$\sim$0 to heavily obscured areas with \Av\ values up to $\sim20-30$\,mag. In most of the objects in the sample ($\sim$70\%), the nucleus of the galaxy coincides with the peak in the extinction maps, with values that range from \Av$\sim3$\,mag up to \Av$\sim$17\,mag. 

We obtained the \Av\ distribution of the individual galaxies on a spaxel-by-spaxel basis (see Fig.~\ref{figure:av_distributions}). The individual \Av\ distributions show a wide range of values with most of them spread between \Av$\sim$1 and \Av$\sim$20 mag, with no clear evidence of any dependence with \LTIR. However, as a class (see Fig.~\ref{figure:samples_distributions}), ULIRGs show \Av\ values (median of 6.5\,mag, mode of $\sim$7-8\,mag) higher than those for LIRGs (median of 5.3\,mag, mode of $\sim$3-4\,mag). The \Av\ distribution in LIRGs shows a mild decrease as a function of galactocentric distances of up to 1\,kpc and a flattening at larger distances (2-3\,kpc). No similar behaviour is detected in ULIRGs, most likely owing to the lower linear resolution of the observations.

To study the effect of the spatial sampling (i.e. physical scale per spaxel) in the derived extinction values at increasing galaxy distances, the individual \Brg\ and \Brd\ maps of our subsample of local LIRGs (at an average distance of 63\,Mpc) have been artificially smeared. These simulations have shown that the spatial resolution plays an important role in shaping the \Av\ distributions. The median value of the visual extinction measured on the LIRG subsample decreases as a function of the linear resolution/distance by a factor $\sim$ 0.8 at the average distance (328\,Mpc, 0.2 kpc/spaxel) of our ULIRG sample, and up to $\sim$0.67 for distances above 800\,Mpc (0.4\,kpc/spaxel). This distance effect would have implications in the derivation of the intrinsic extinction, and subsequent properties, such as SFR, \Si, and the KS-law, in high-z star-forming galaxies, even in AO-based spectroscopy. If local LIRGs are analogues of the main-sequence star-forming galaxies at cosmological distances, the extinction values (\Av) derived from the observed emission lines in these high-z sources would need to be increased by a factor 1.4 on average.

\begin{acknowledgements}
We thank the anonymous referee for useful comments and suggestions that improved the final content of this paper.
This work was supported by the Spanish Ministry of Science and Innovation (MICINN) under grants BES-2008-007516, ESP2007-65475-C02-01, and AYA2010-21161-C02-01.
This paper made use of the plotting package \emph{jmaplot}, developed by Jes\'us Ma\'iz-Apell\'aniz \url{http://jmaiz.iaa.es/software/jmaplot/current/html/jmaplot_overview.html}.
Based on observations collected at the European Organisation for Astronomical Research in the Southern Hemisphere, Chile, programmes 077.B-0151A, 078.B-0066A, and 081.B-0042A.
This research made use of the NASA/IPAC Extragalactic Database (NED), which is operated by the Jet Propulsion Laboratory, California Institute of Technology, under contract with the National Aeronautics and Space Administration. 
\end{acknowledgements}

\clearpage
\setcounter{figure}{0}
\setcounter{fake_fig}{\value{figure}}
\refstepcounter{fake_fig}
\label{figure:LIRG}
\renewcommand{\thefigure}{\arabic{figure}\alph{subfig}}
\setcounter{subfig}{1}

\begin{figure*}[t]
\begin{center}
\begin{tabular}{c}
\includegraphics[angle=0, width=.86\textwidth]{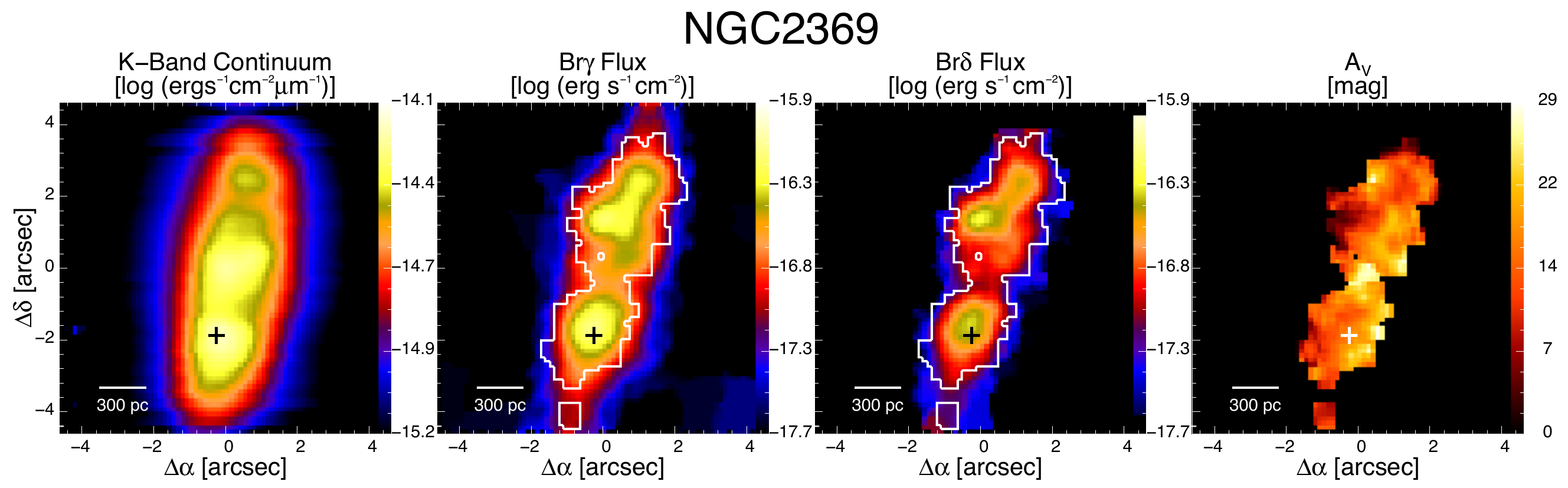} \\
\includegraphics[angle=0, width=.86\textwidth]{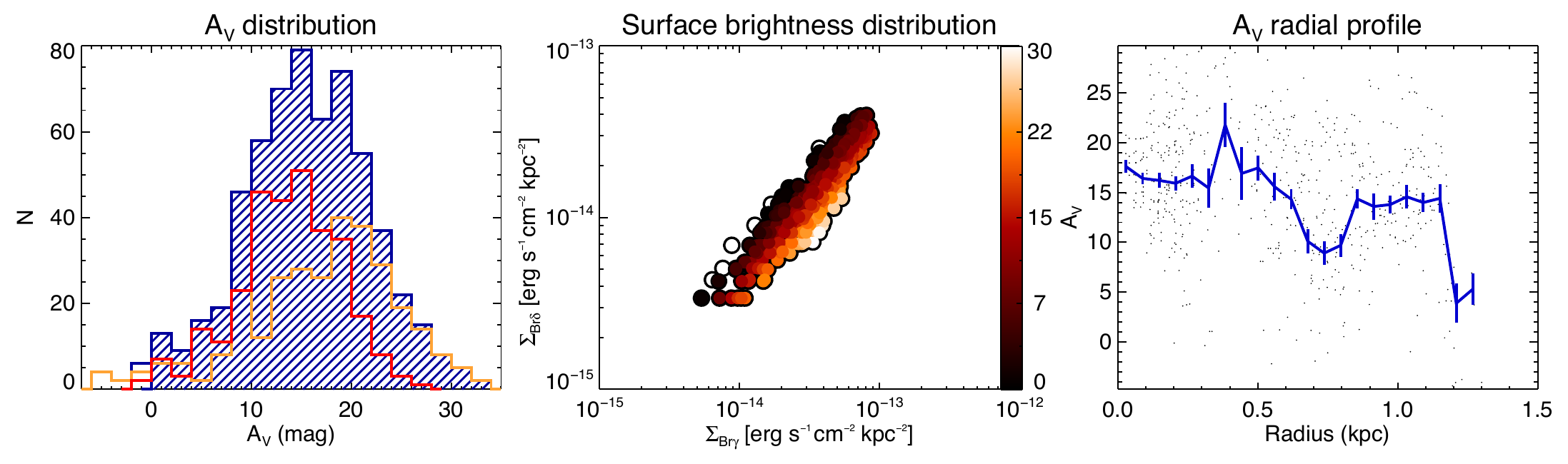} \\
\end{tabular}
\end{center}
\caption{\tiny \object{NGC 2369}. Top panels show the SINFONI K band continuum emission, the observed maps (not corrected from extinction) of the lines \Brgl\ and \Brdl, together with the \Av\ map. The white contour englobes those spaxels above $\rm S/N=4$ considered to build the \Av\ map and distribution. The brightest spaxel of the K band continuum is marked with a plus symbol (+), and has been used as reference to obtain the radial profile in the bottom right panel. The secondary nucleus, if present, is marked with a cross ($\times$). Bottom left panel shows the the Av distribution of all valid spaxels (blue histogram) and the distributions of the those spaxels above (red) and below (yellow) the median value of the \Brd\ S/N distribution. The relationship between the surface brightness of the lines and the \Av\ is shown in the central panel, where the points with \Av$<$0 are outlined with a black contour. Finally, the radial distribution of the extinction is shown in the bottom right panel, where the blue line represents the mean value of \Av\ for different radial bins and its error. The bins are obtained as the 1/30 of the total radial coverage of the map.}
\label{figure:NGC2369}
\end{figure*}

\addtocounter{figure}{-1}
\addtocounter{subfig}{1}
\begin{figure*}[!hb]
\begin{center}
\begin{tabular}{c}
\includegraphics[angle=0, width=.86\textwidth]{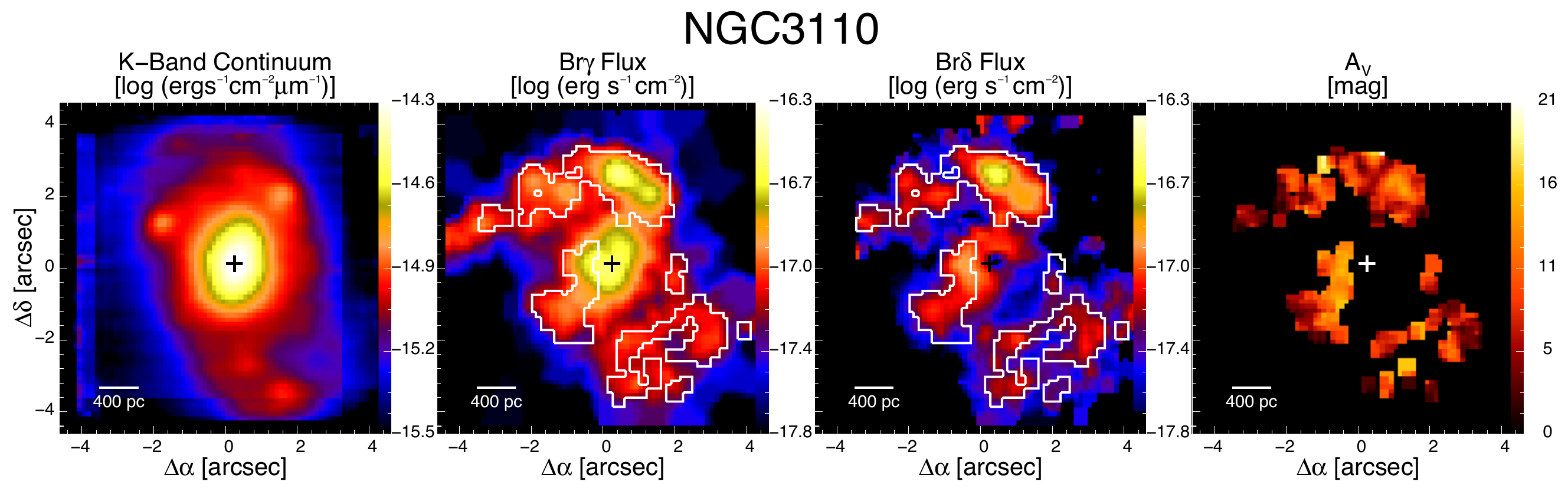} \\
\includegraphics[angle=0, width=.86\textwidth]{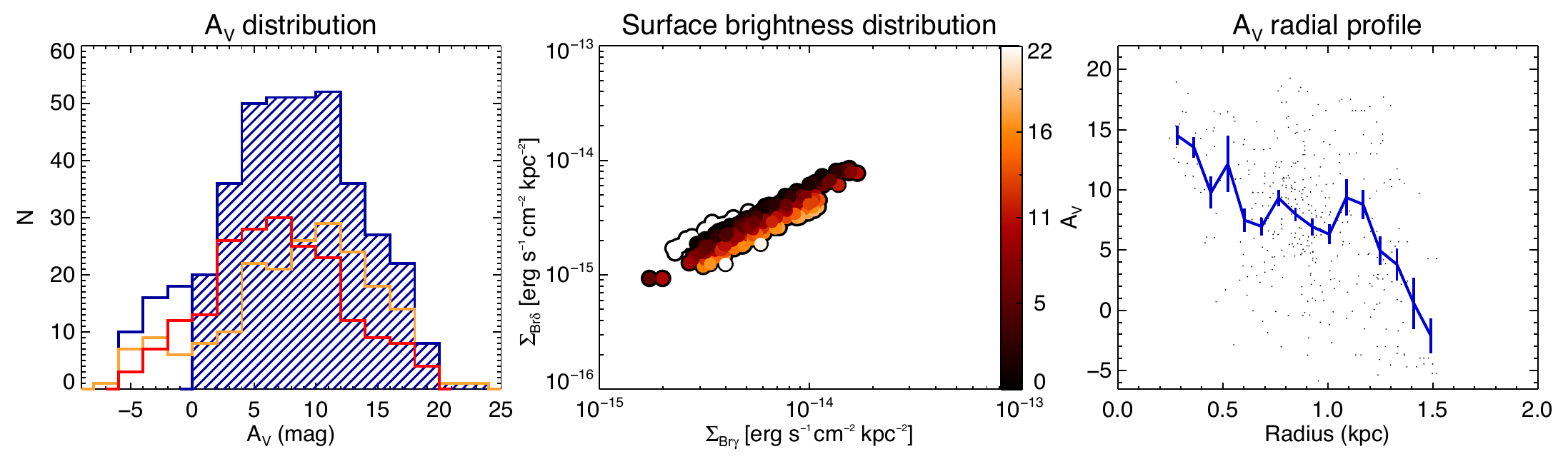} \\
\end{tabular}
\end{center}
\caption{\object{NGC 3110}. Same as Fig.~\ref{figure:NGC2369} but for \object{NGC 3110}}
\label{figure:NGC3110}
\end{figure*}

\addtocounter{figure}{-1}
\addtocounter{subfig}{1}
\begin{figure*}[t]
\begin{center}
\begin{tabular}{c}
\includegraphics[angle=0, width=0.99\textwidth]{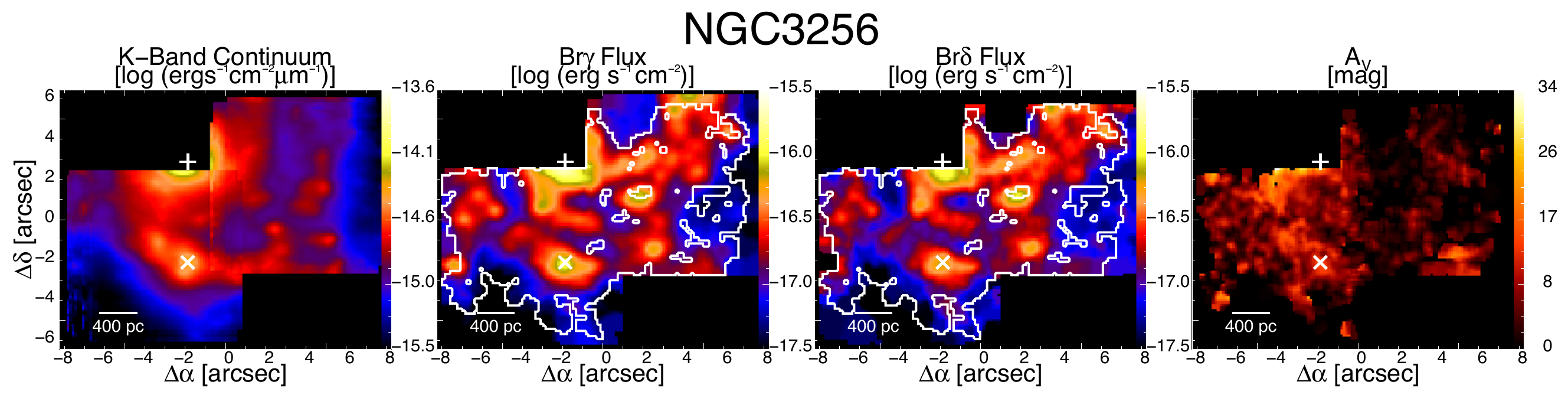} \\
\includegraphics[angle=0, width=.9\textwidth]{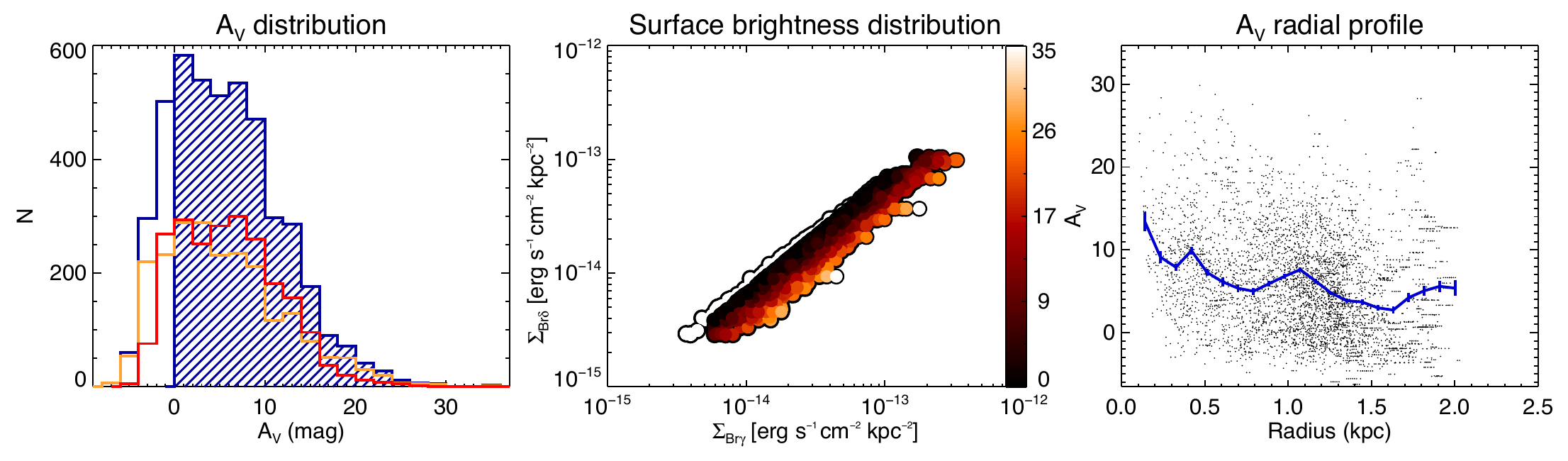} \\
\end{tabular}
\end{center}
\caption{\object{NGC 3256}. Same as Fig.~\ref{figure:NGC2369} but for \object{NGC 3256}. Please note that the central spaxel lays outside the FoV since the nucleus was not observed in K-band. See text and Paper I for further details.}
\label{figure:NGC3256}
\end{figure*}

\addtocounter{figure}{-1}
\addtocounter{subfig}{1}
\begin{figure*}[!hb]
\begin{center}
\begin{tabular}{c}
\includegraphics[angle=0, width=.98\textwidth]{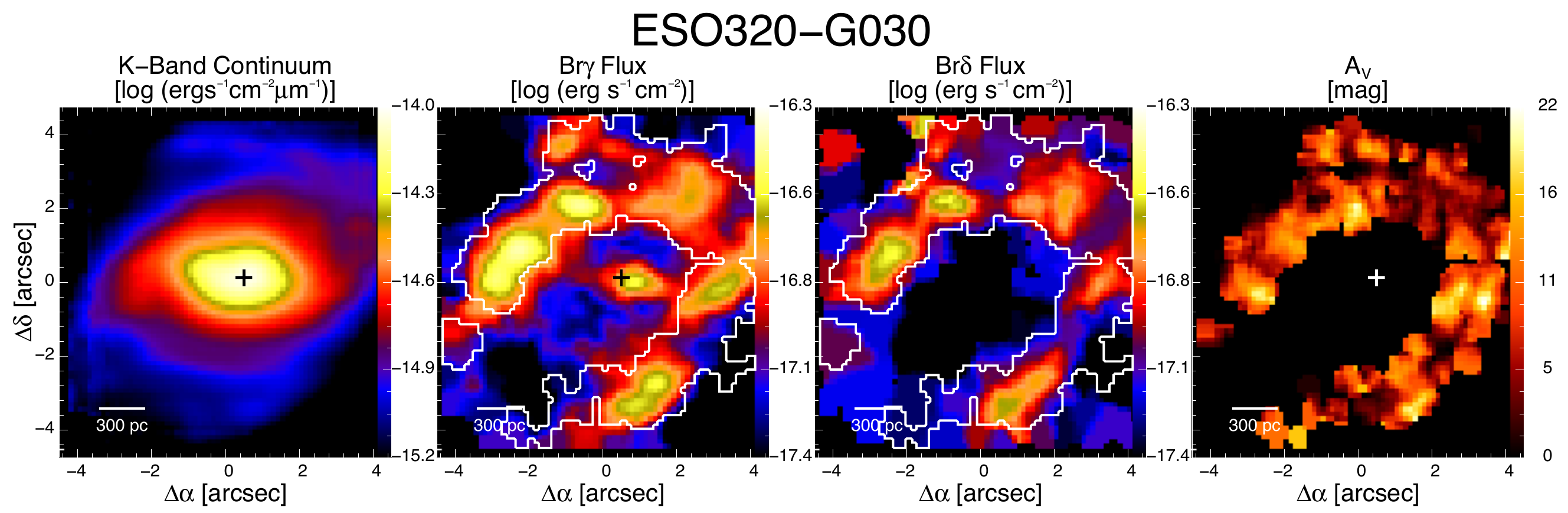} \\
\includegraphics[angle=0, width=.92\textwidth]{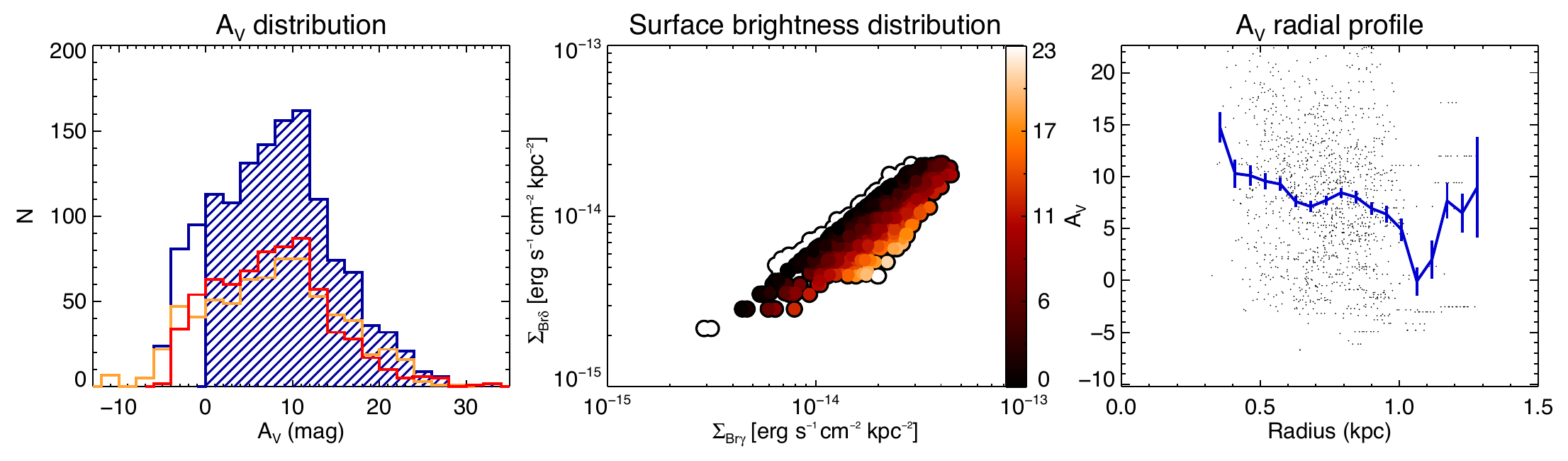} \\
\end{tabular}
\end{center}
\caption{\object{ESO 320-G030}. Same as Fig.~\ref{figure:NGC2369} but for \object{ESO 320-G030}}
\label{figure:ESO320}
\end{figure*}

\addtocounter{figure}{-1}
\addtocounter{subfig}{1}
\begin{figure*}[t]
\begin{center}
\begin{tabular}{c}
\includegraphics[angle=0, width=.98\textwidth]{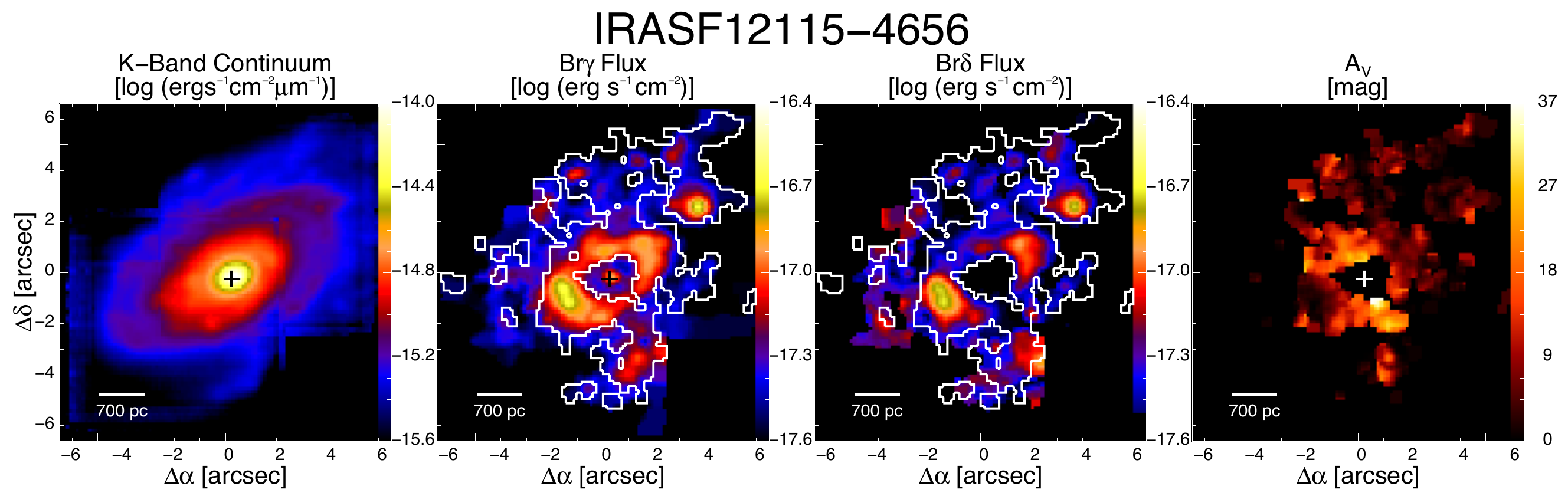} \\
\includegraphics[angle=0, width=.9\textwidth]{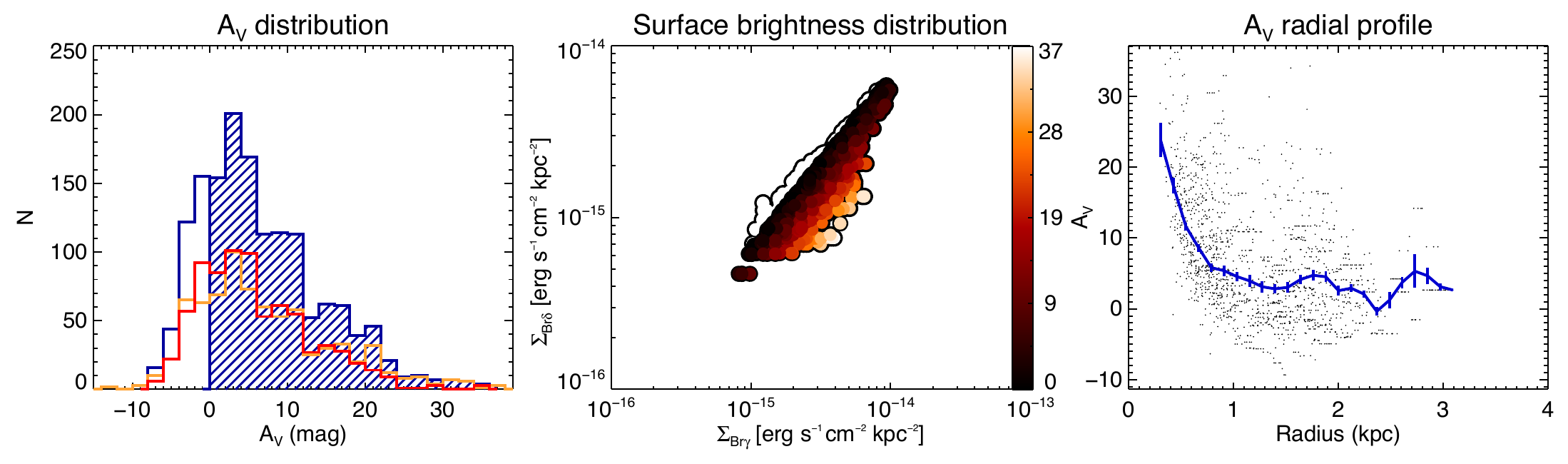} \\
\end{tabular}
\end{center}
\caption{\object{IRASF 12115-4656}. Same as Fig.~\ref{figure:NGC2369} but for \object{IRASF 12115-4656}}
\label{figure:IRASF12115}
\end{figure*}

\addtocounter{figure}{-1}
\addtocounter{subfig}{1}
\begin{figure*}[!hb]
\begin{center}
\begin{tabular}{c}
\includegraphics[angle=0, width=.98\textwidth]{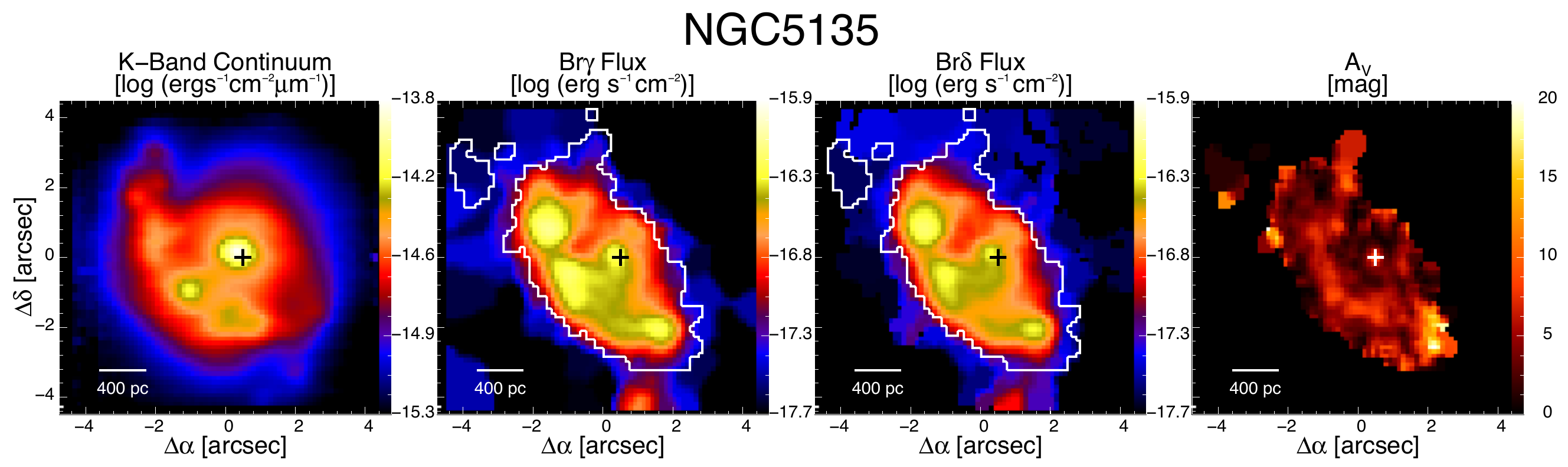} \\
\includegraphics[angle=0, width=.9\textwidth]{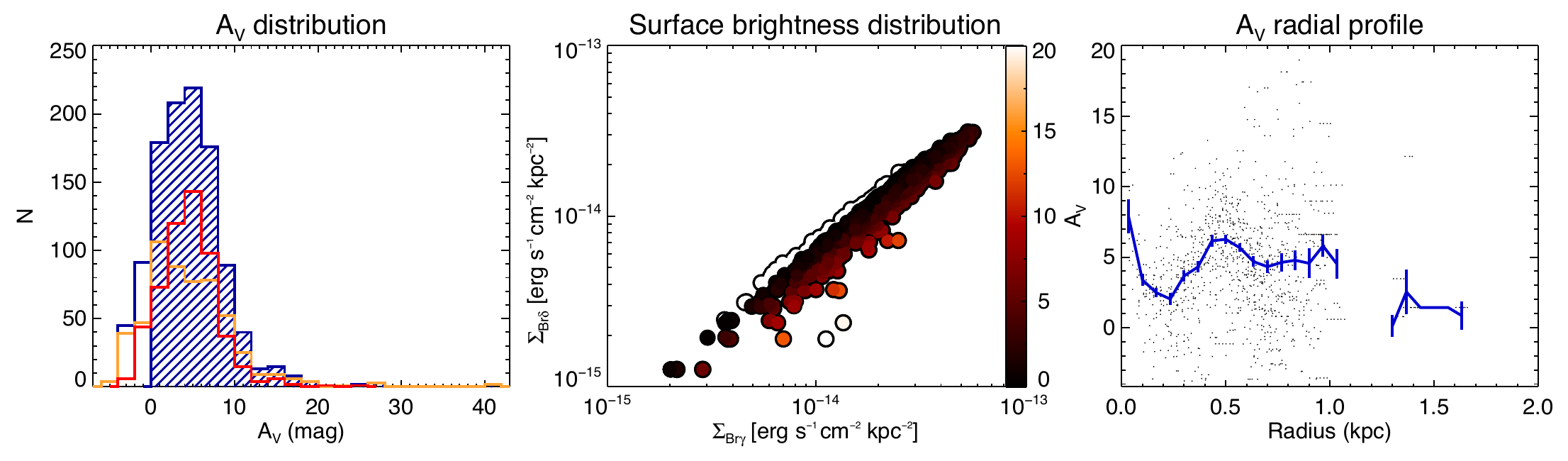} \\
\end{tabular}
\end{center}
\caption{\object{NGC 5135}. Same as Fig.~\ref{figure:NGC2369} but for \object{NGC 5135}}
\label{figure:NGC5135}
\end{figure*}

\addtocounter{figure}{-1}
\addtocounter{subfig}{1}
\begin{figure*}[t]
\begin{center}
\begin{tabular}{c}
\includegraphics[angle=0, width=.98\textwidth]{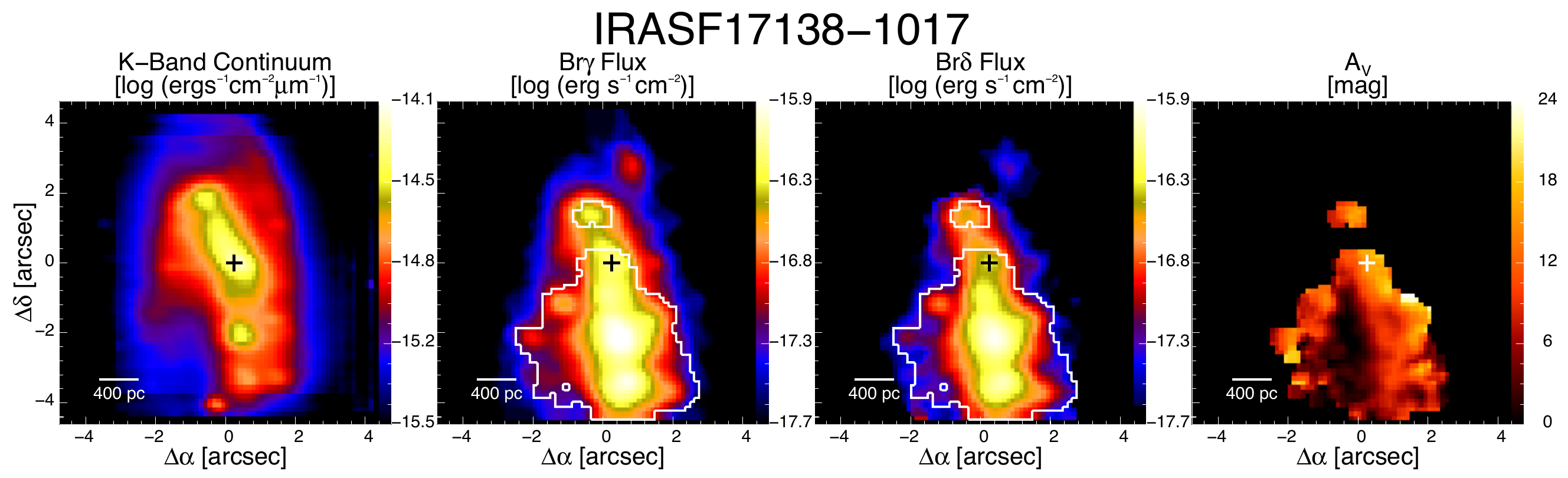} \\
\includegraphics[angle=0, width=.9\textwidth]{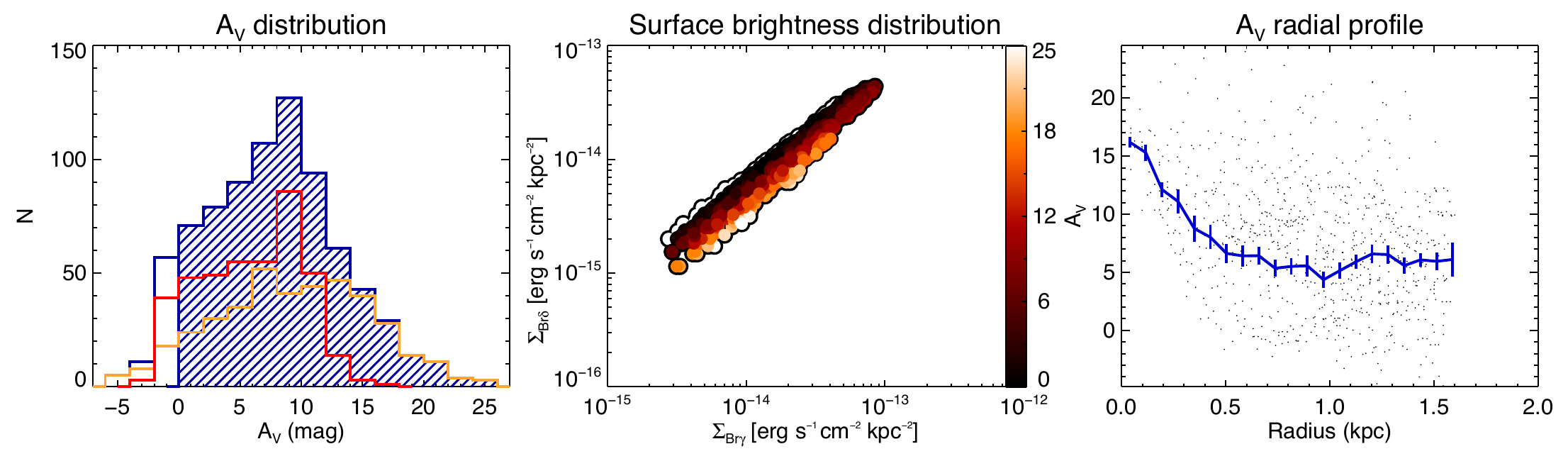} \\
\end{tabular}
\end{center}
\caption{\object{IRASF 17138-1017}. Same as Fig.~\ref{figure:NGC2369} but for \object{IRASF 17138-1017}}
\label{figure:IRASF17138}
\end{figure*}

\addtocounter{figure}{-1}
\addtocounter{subfig}{1}
\begin{figure*}[!hb]
\begin{center}
\begin{tabular}{c}
\includegraphics[angle=0, width=.98\textwidth]{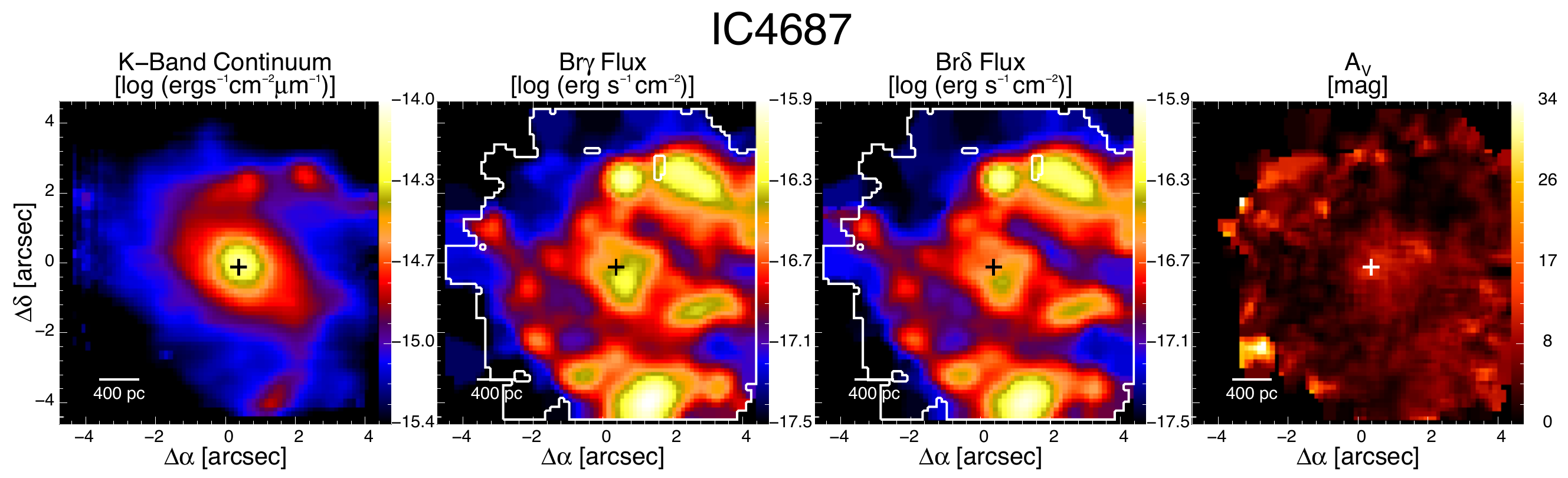} \\
\includegraphics[angle=0, width=.9\textwidth]{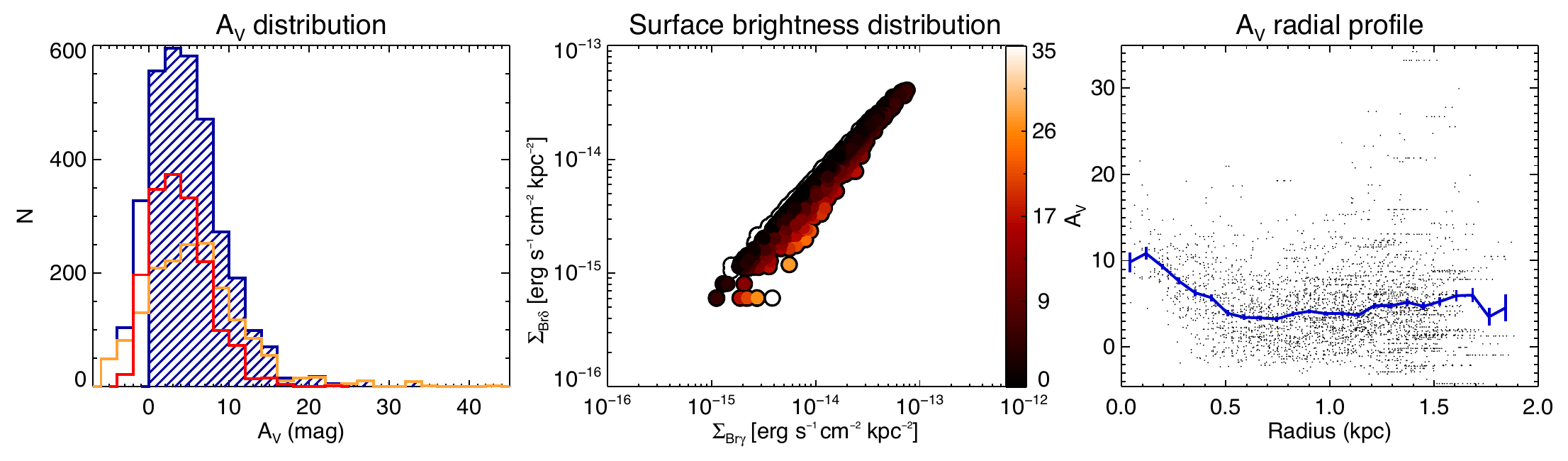} \\
\end{tabular}
\end{center}
\caption{\object{IC 4687}. Same as Fig.~\ref{figure:NGC2369} but for \object{IC 4687}}
\label{figure:IC4687}
\end{figure*}

\addtocounter{figure}{-1}
\addtocounter{subfig}{1}
\begin{figure*}[t]
\begin{center}
\begin{tabular}{c}
\includegraphics[angle=0, width=.81\textwidth]{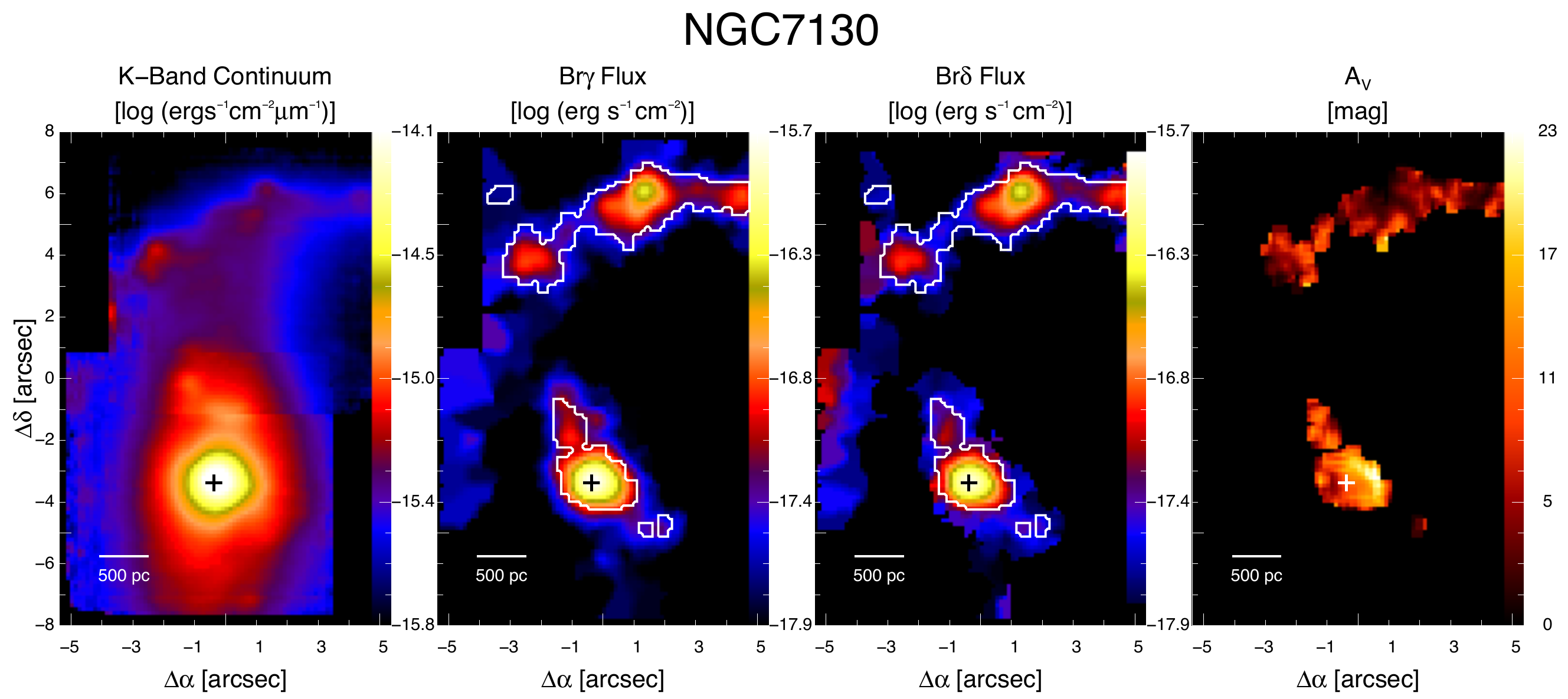} \\
\includegraphics[angle=0, width=.88\textwidth]{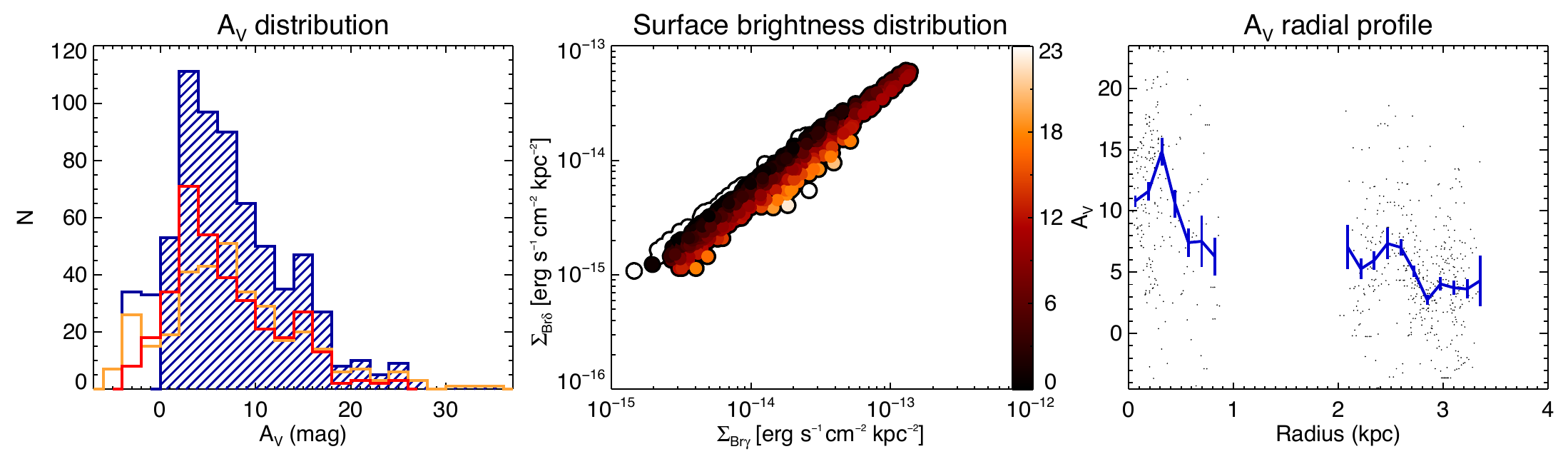} \\
\end{tabular}
\end{center}
\caption{\object{NGC 7130}. Same as Fig.~\ref{figure:NGC2369} but for \object{NGC 7130}}
\label{figure:NGC7130}
\end{figure*}

\addtocounter{figure}{-1}
\addtocounter{subfig}{1}
\begin{figure*}[!hb]
\begin{center}
\begin{tabular}{c}
\includegraphics[angle=0, width=.87\textwidth]{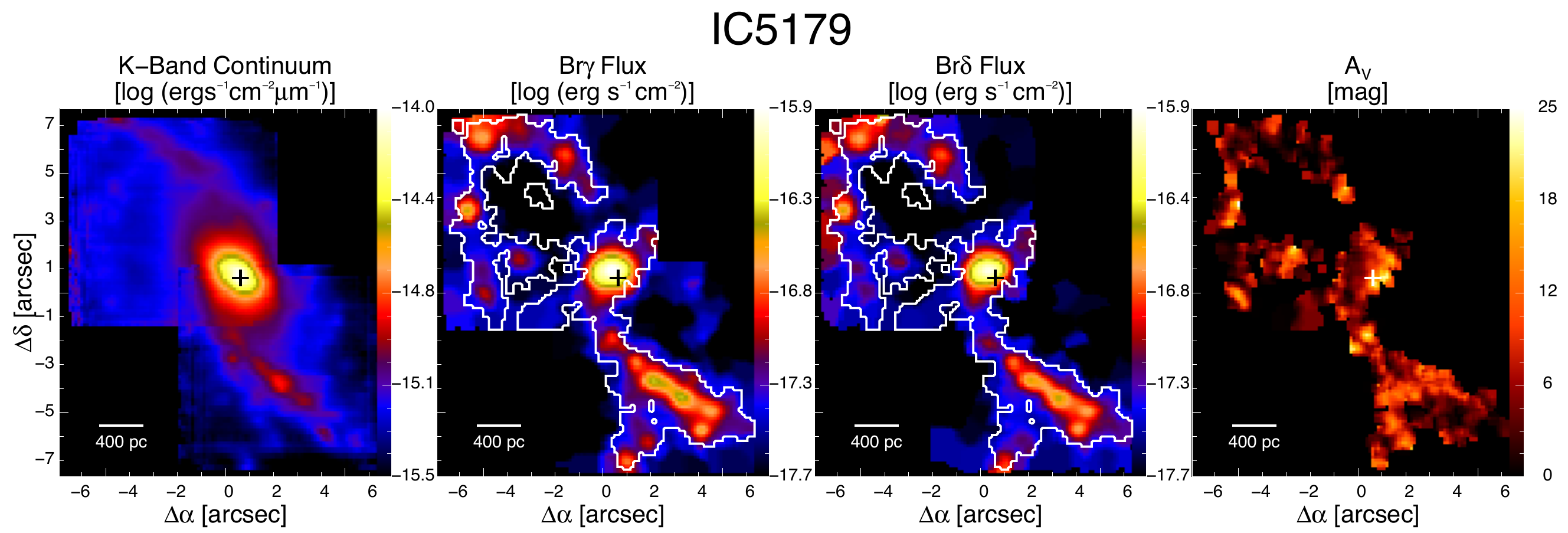} \\
\includegraphics[angle=0, width=.86\textwidth]{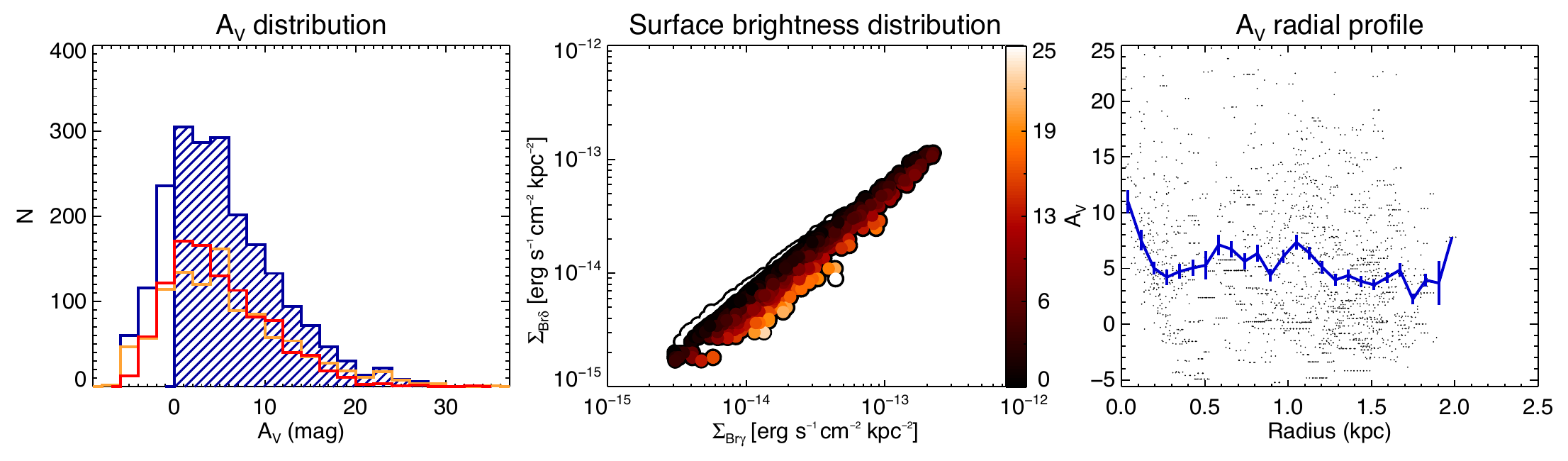} \\
\end{tabular}
\end{center}
\caption{\object{IC 5179}. Same as Fig.~\ref{figure:NGC2369} but for \object{IC 5179}}
\label{figure:IC5179}
\end{figure*}

\refstepcounter{fake_fig}
\label{figure:ULIRG}

\addtocounter{figure}{0}
\setcounter{subfig}{1}
\begin{figure*}[t]
\begin{center}
\begin{tabular}{c}
\includegraphics[angle=0, width=.86\textwidth]{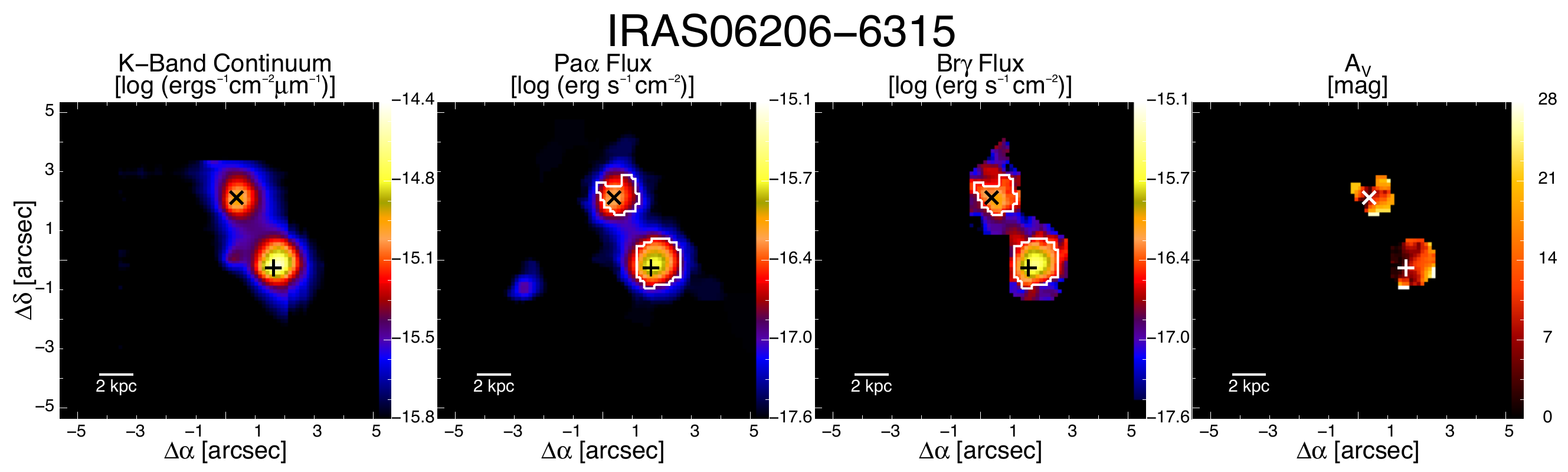} \\
\includegraphics[angle=0, width=.80\textwidth]{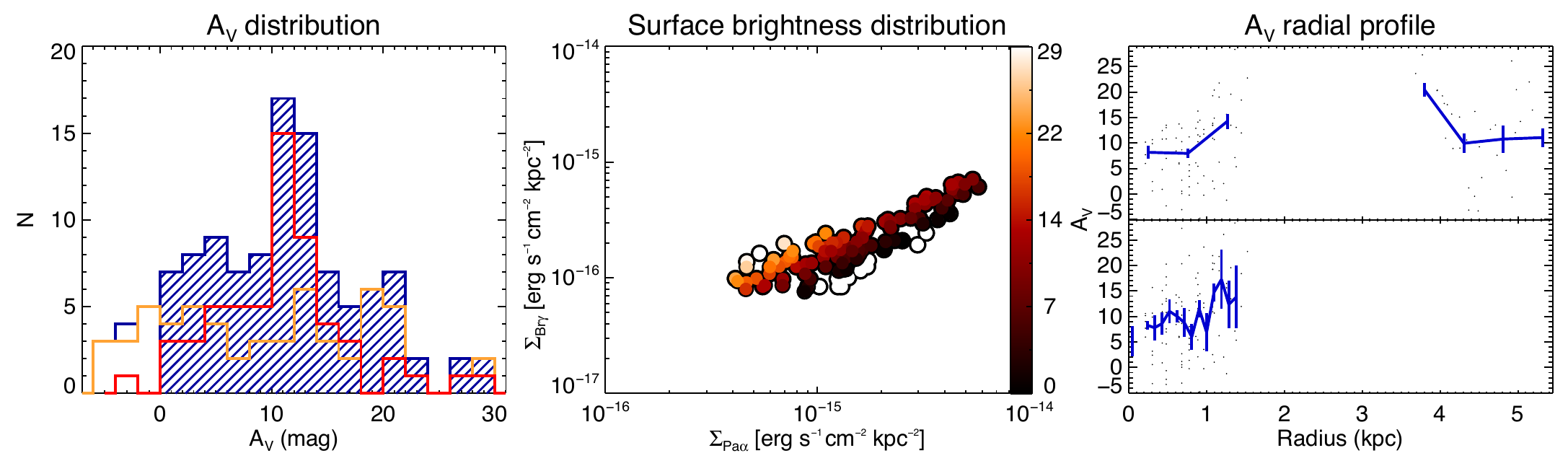} \\
\end{tabular}
\end{center}
\caption{\tiny \object{IRAS 06206-6315}. Top panels show the SINFONI K band continuum emission, the observed maps (not corrected from extinction) of the lines \Pal\ and \Brgl, together with the \Av\ map. The white contour englobes those spaxels above $\rm S/N=4$ considered to build the \Av\ map and distribution. The brightest spaxel of the K band continuum is marked with a plus symbol (+), and has been used as reference to obtain the radial profile in the bottom right panel. The secondary nucleus, if present, is marked with a cross ($\times$). Bottom left panel shows the Av distribution of all valid spaxels (blue histogram) and the distributions of the those spaxels above (red) and below (yellow) the median value of the \Brg\ S/N distribution. The relationship between the surface brightness of the lines and the \Av\ is shown in the central panel, where the points with \Av$<$0 are outlined with a black contour. Finally, the radial distribution of the extinction is shown in the bottom right panel, where the blue line represents the mean value of \Av\ for different radial bins and its error. The bins are obtained as the 1/30 of the total radial coverage of the map. For those objects with multiple components, the top inset shows the \Av\ radial profile of the system taking the main nucleus as the centre, whereas the bottom subpanel shows the radial profile obtained by extracting the profiles of each component separately and plotting them in the same reference frame.}
\label{figure:IRAS06206}
\end{figure*}

\addtocounter{figure}{-1}
\addtocounter{subfig}{1}
\begin{figure*}[!hb]
\begin{center}
\begin{tabular}{c}
\includegraphics[angle=0, width=.86\textwidth]{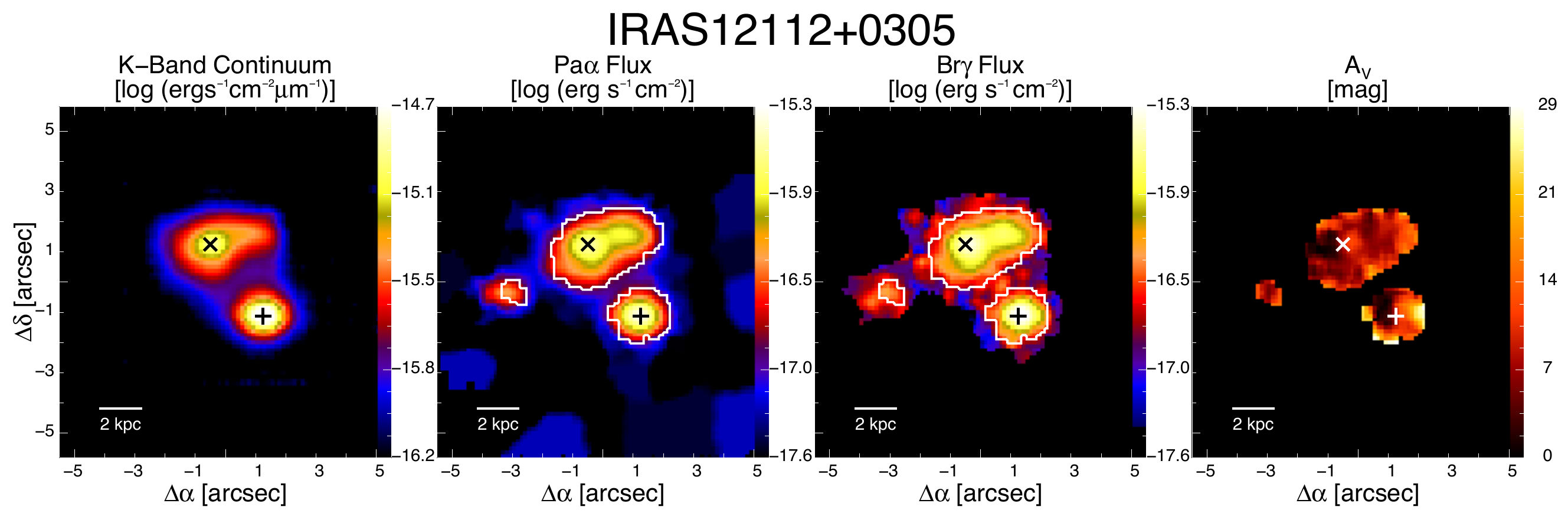} \\
\includegraphics[angle=0, width=.80\textwidth]{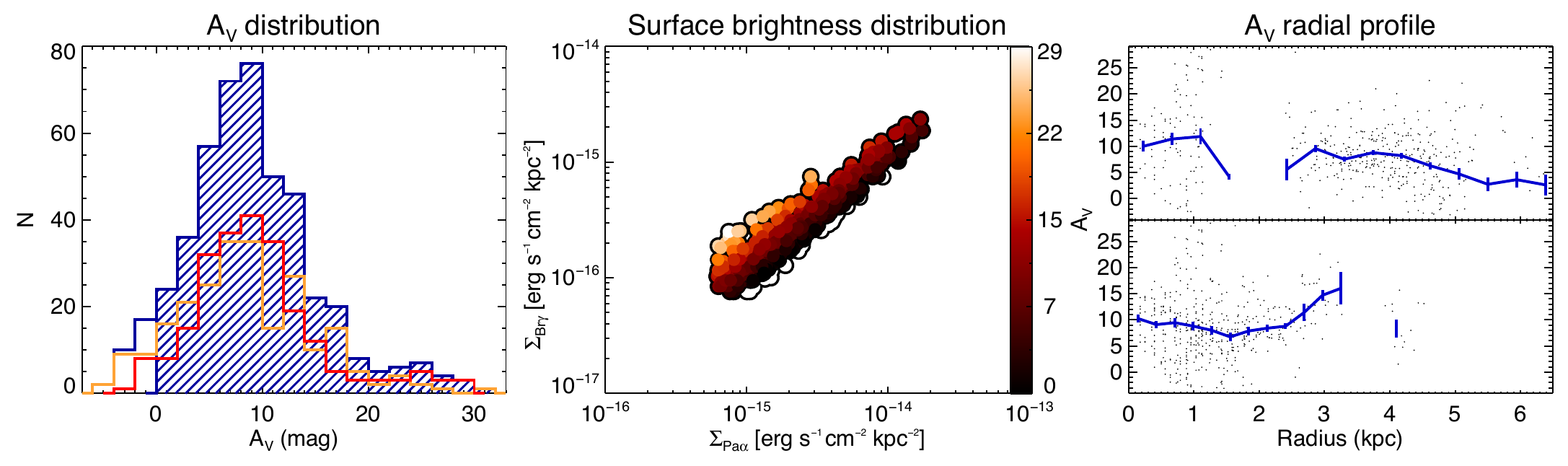} \\
\end{tabular}
\end{center}
\caption{\object{IRAS 12112+0305}. Same as Fig. \ref{figure:IRAS06206} but for \object{IRAS 12112+0305}.}
\label{figure:IRAS12112}
\end{figure*}

\addtocounter{figure}{-1}
\addtocounter{subfig}{1}
\begin{figure*}[t]
\begin{center}
\begin{tabular}{c}
\includegraphics[angle=0, width=.97\textwidth]{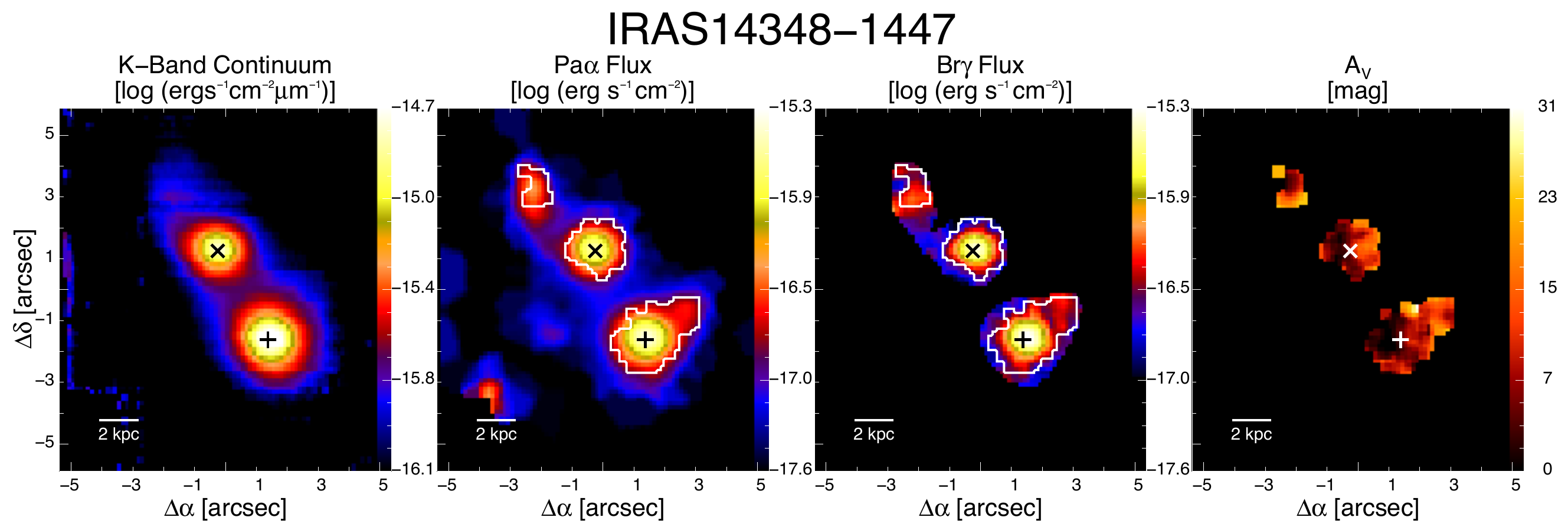} \\
\includegraphics[angle=0, width=.9\textwidth]{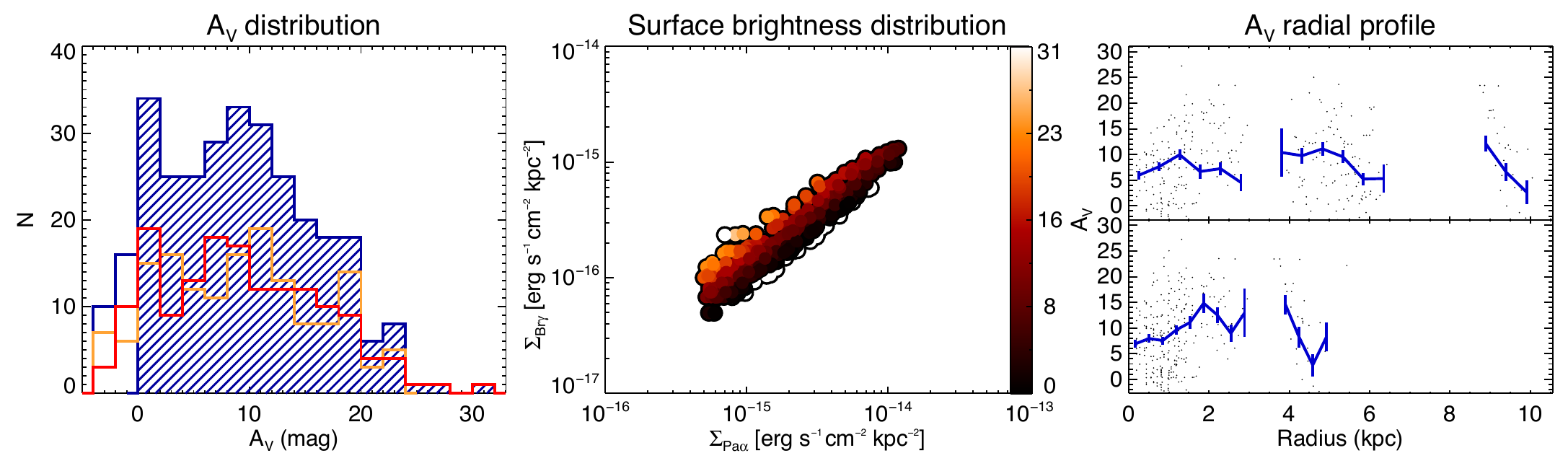} \\
\end{tabular}
\end{center}
\caption{\object{IRAS 14348-1447}. Same as Fig. \ref{figure:IRAS06206} but for \object{IRAS 14348-1447}.}
\label{figure:IRAS14348}
\end{figure*}

\addtocounter{figure}{-1}
\addtocounter{subfig}{1}
\begin{figure*}[!hb]
\begin{center}
\begin{tabular}{c}
\includegraphics[angle=0, width=.98\textwidth]{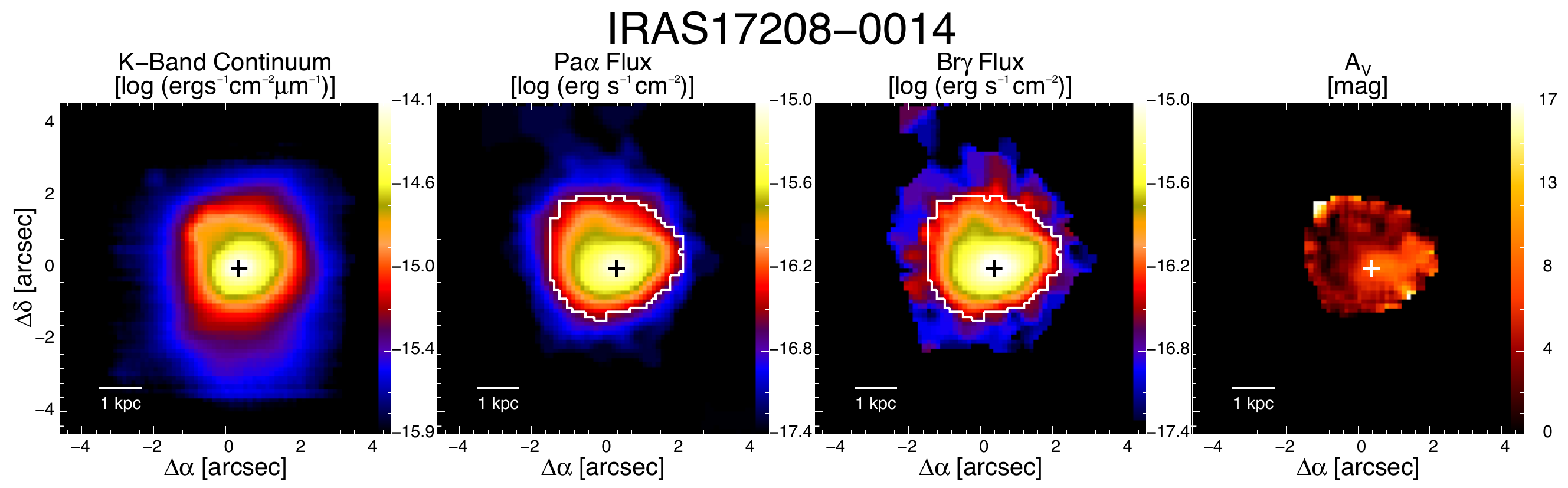} \\
\includegraphics[angle=0, width=.9\textwidth]{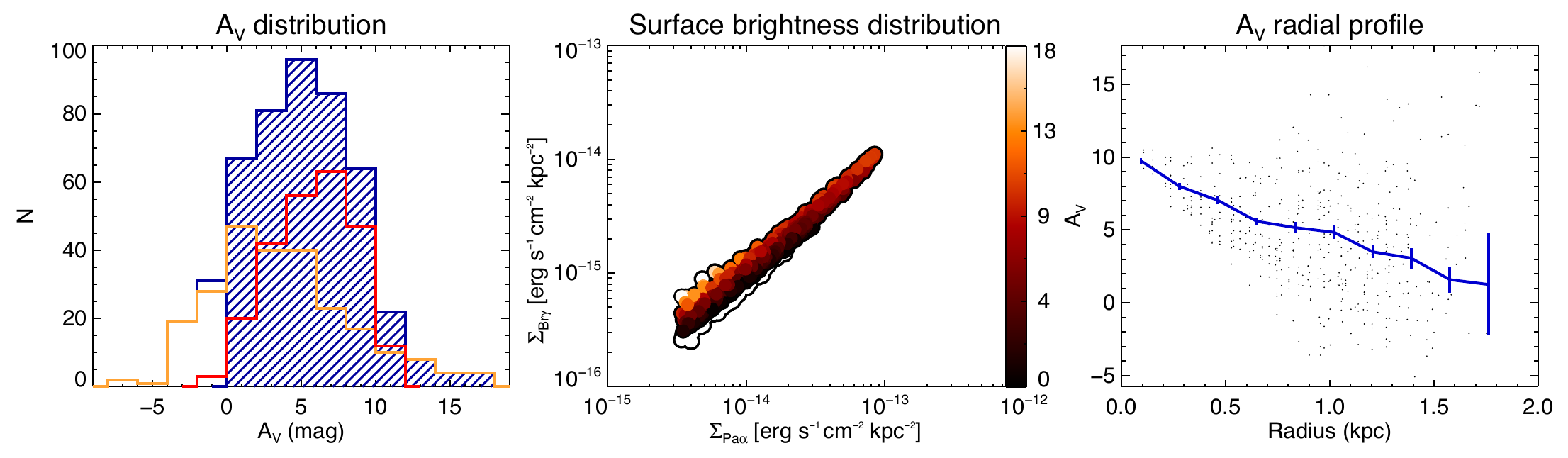} \\
\end{tabular}
\end{center}
\caption{\object{IRAS 17208-0014}. Same as Fig. \ref{figure:IRAS06206} but for \object{IRAS 17208-0014}.}
\label{figure:IRAS17208}
\end{figure*}

\addtocounter{figure}{-1}
\addtocounter{subfig}{1}
\begin{figure*}[t]
\begin{center}
\begin{tabular}{c}
\includegraphics[angle=0, width=.98\textwidth]{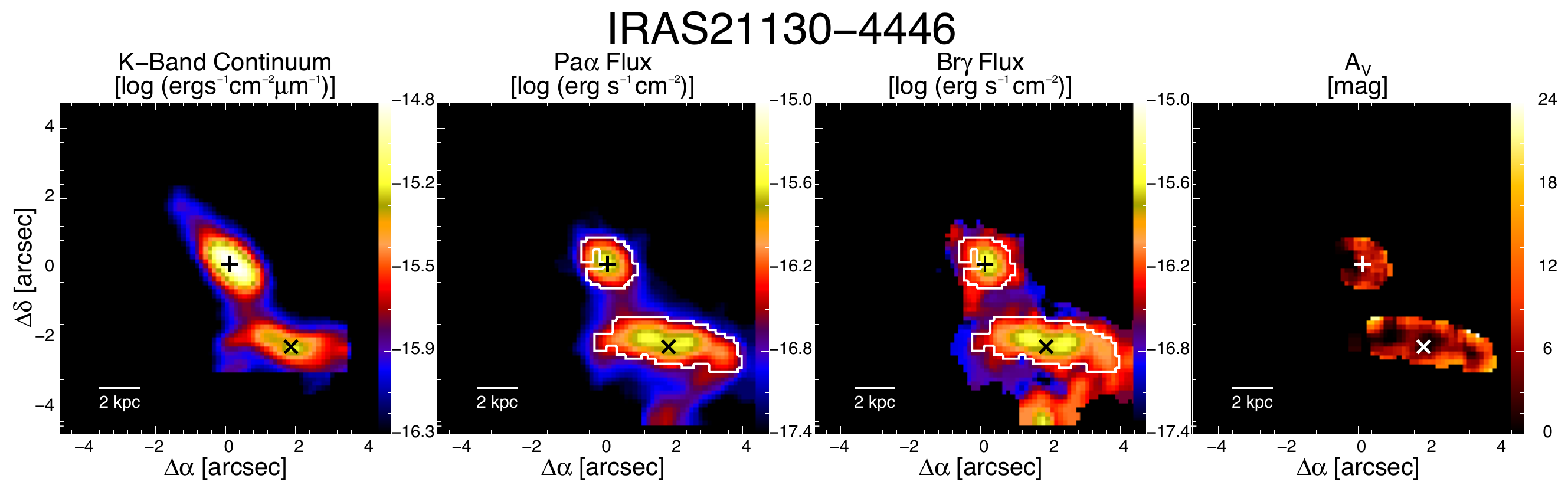} \\
\includegraphics[angle=0, width=.9\textwidth]{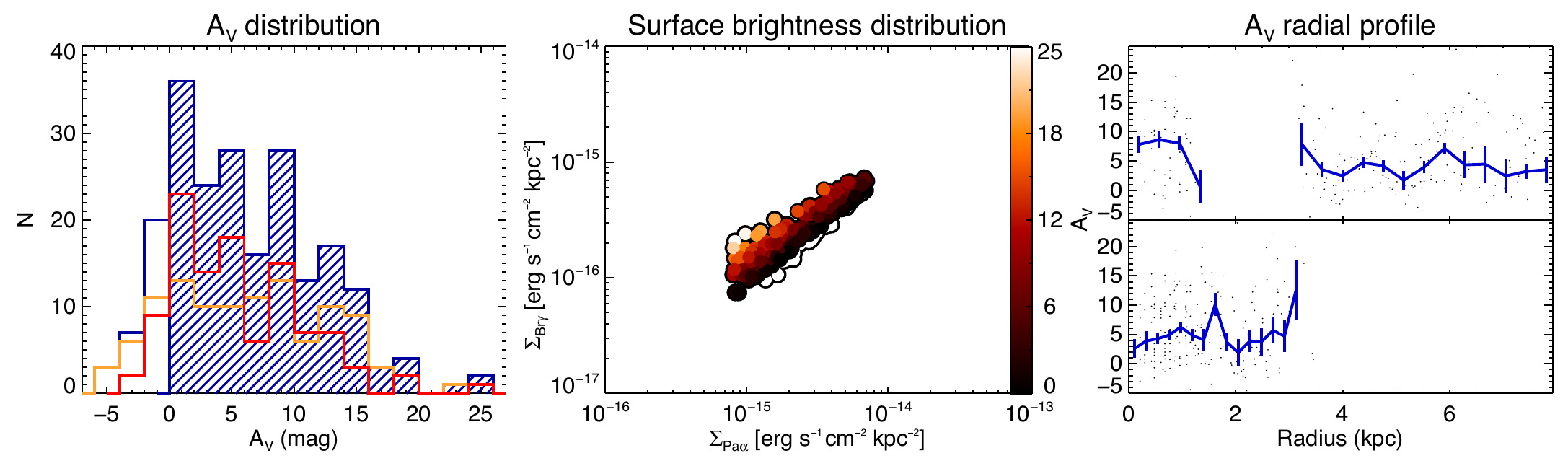} \\
\end{tabular}
\end{center}
\caption{\object{IRAS 21130-4446}. Same as Fig. \ref{figure:IRAS06206} but for \object{IRAS 21130-4446}.}
\label{figure:IRAS21130}
\end{figure*}

\addtocounter{figure}{-1}
\addtocounter{subfig}{1}
\begin{figure*}[!hb]
\begin{center}
\begin{tabular}{c}
\includegraphics[angle=0, width=.98\textwidth]{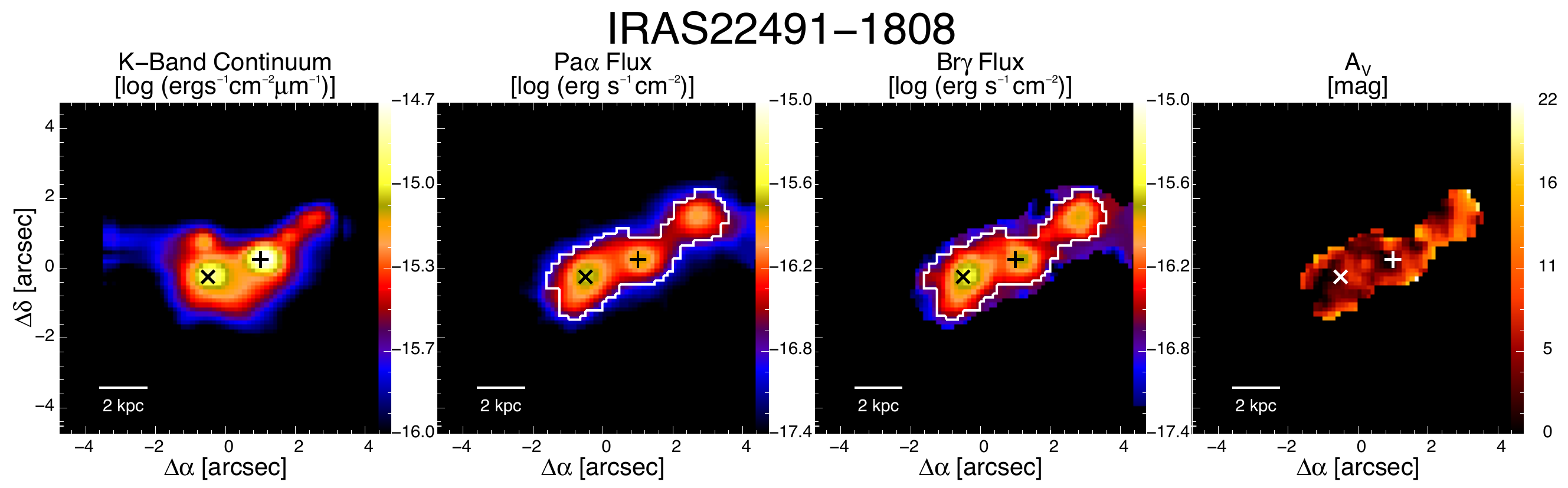} \\
\includegraphics[angle=0, width=.9\textwidth]{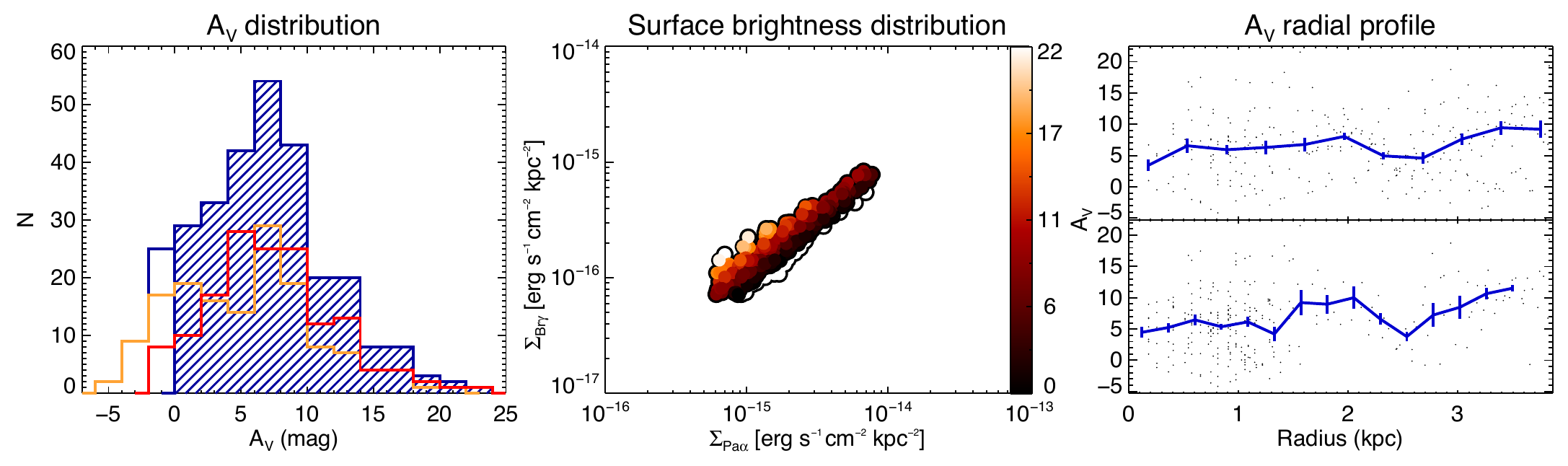} \\
\end{tabular}
\end{center}
\caption{\object{IRAS 22491-1808}. Same as Fig. \ref{figure:IRAS06206} but for \object{IRAS 22491-1808}.}
\label{figure:IRAS22491}
\end{figure*}

\addtocounter{figure}{-1}
\addtocounter{subfig}{1}
\begin{figure*}[t]
\begin{center}
\begin{tabular}{c}
\includegraphics[angle=0, width=.98\textwidth]{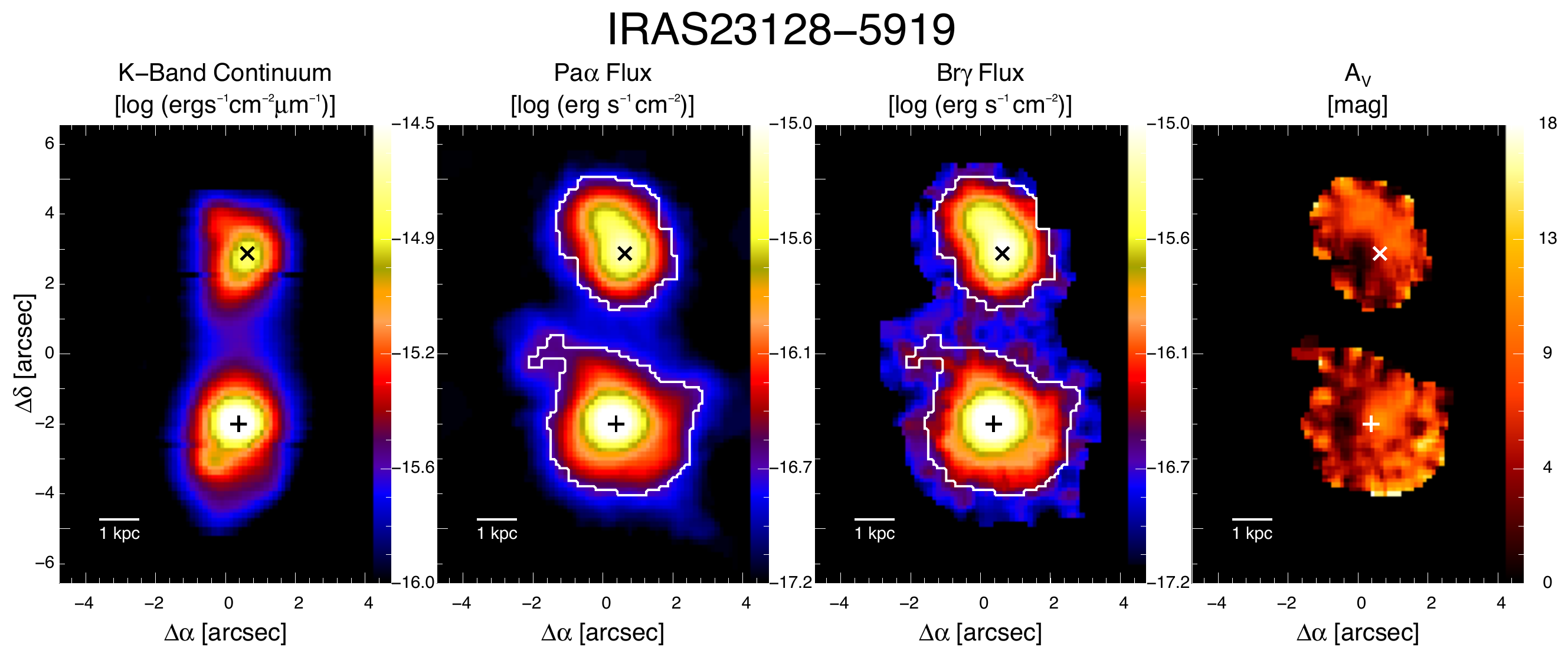} \\
\includegraphics[angle=0, width=.95\textwidth]{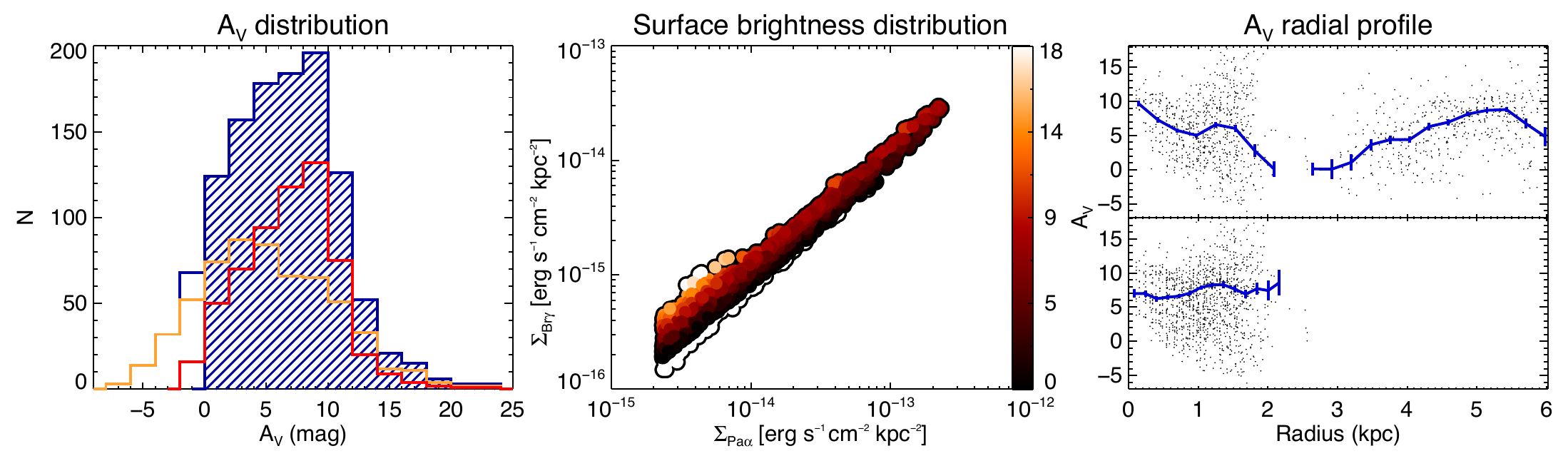} \\
\end{tabular}
\end{center}
\caption{\object{IRAS 23128-5919}. Same as Fig. \ref{figure:IRAS06206} but for \object{IRAS 23128-5919}.}
\label{figure:IRAS23128}
\end{figure*}

\bibliographystyle{aa}
\bibliography{bib_file}
\end{document}